\newcommand{\editx}[1]{\textcolor{red}{[#1]}}

\documentclass{article}

\usepackage{arxiv}
\usepackage{titlesec}

\titleformat{\paragraph}
{\normalfont\normalsize\bfseries}{\theparagraph}{1em}{}
\titlespacing*{\paragraph}
{0pt}{1.25ex plus 1ex minus .2ex}{1.0ex plus .2ex}

\usepackage{xcolor}
\usepackage[utf8]{inputenc} 
\usepackage[T1]{fontenc}    
\usepackage{hyperref}       
\usepackage{url}            
\usepackage{booktabs}       
\usepackage{amsfonts}       
\usepackage{nicefrac}       
\usepackage{microtype}      
\usepackage{lipsum}		
\usepackage{graphicx}
\usepackage{fancybox}
\usepackage{natbib}
\usepackage{doi}
\usepackage{enumitem}
\usepackage{siunitx}
 \usepackage{float}

\title{Hyperspectral Image Restoration and Super-resolution with Physics-Aware Deep Learning for Biomedical Applications}

\author{ 
	\href{}{\hspace{1mm}Yuchen Xiang*$^{,1}$} \\
	\texttt{yuchen.xiang11@imperial.ac.uk}\\
    	\And
    \href{}{\hspace{1mm}Zhaolu Liu*$^{,2}$} \\
	\texttt{zhaolu.liu16@imperial.ac.uk} \\
	\And
    \href{}{\hspace{1mm}Monica Emili Garcia-Segura$^{1,4}$}\\
    \texttt{me485@cam.ac.uk}\\
    \And
    \href{}{\hspace{1mm}Daniel Simon$^{1}$}\\
	\texttt{d.simon19@imperial.ac.uk}\\
    \And
    \href{}{\hspace{1mm}Boxuan Cao$^{1}$}\\
	\texttt{caoboxuan1998@gmail.com}\\
    \And
    \href{}{\hspace{1mm}Vincen Wu$^{1}$}\\
	\texttt{vwu@ethz.ch}\\
    \And
    \href{}{\hspace{1mm}Kenneth Robinson$^{1}$}\\
	\texttt{k.robinson@imperial.ac.uk}\\
    \And
    \href{}{\hspace{1mm}Yu Wang$^{3}$}\\
	\texttt{yu.wang1@imperial.ac.uk}\\
    \And
    \href{}{\hspace{1mm}Ronan Battle$^{3}$}\\
	\texttt{ronan.battle15@imperial.ac.uk}\\
    \And
   \href{}{\hspace{1mm}Najah Sobhan$^{7}$}\\
	\texttt{m.najah@imperial.ac.uk}\\
    \And
    \href{}{\hspace{1mm}Robert T. Murray$^{3}$}\\
	\texttt{robert.murray10@imperial.ac.uk}\\
    \And
    \href{}{\hspace{1mm} Xavier Altafaj$^{5,6}$}\\
    \texttt{xaltafaj@ub.edu}\\
    \And
    \href{}{\hspace{1mm} John Marshall$^{7}$}\\
    \texttt{j.f.marshall@qmul.ac.uk}\\
    \And
    \href{}{\hspace{1mm} Luca Peruzzotti-Jametti$^{1,4}$}\\
    \texttt{lp429@cam.ac.uk}\\
    \And
	\href{}{\hspace{1mm}Zoltan Takats$^{1}$}\\
	\texttt{z.takats@imperial.ac.uk}\\\\  
    \href{}{1. Department of Metabolism, Digestion and Reproduction,
	Imperial College London,
	London, UK}\\
   \href{}{2. Department of Mathematics,
	Imperial College London,
	London, UK}\\
     \href{}{3. Department of Physics,
	Imperial College London,
	London, UK}\\
    \href{}{4. Department of Clinical Neurosciences and NIHR Biomedical Research Centre,}\\
    \href{}{University of Cambridge,
    Cambridge, UK}\\
    \href{}{5. Department of Biomedicine, School of Medicine and Health Sciences,}\\
    \href{}{Institute of Neurosciences, University of Barcelona, Barcelona 08036, Spain}\\
    \href{}{6. Agustí Pi i Sunyer Biomedical Research Institute (IDIBAPS), University of Barcelona,}\\ 
    \href{}{Barcelona 08036, Spain}\\
    \href{}{7. Barts Cancer Institute, Cancer Research UK Centre of Excellence,}\\
    \href{}{Queen Mary University of London, London, UK}\\
    \href{}{*these authors contributed equally to the manuscript.}
}

\date{}

\renewcommand{\shorttitle}{HyReS}

\begin{document}
\maketitle

\begin{abstract}
Hyperspectral imaging is a powerful bioimaging tool which can uncover novel insights, thanks to its sensitivity to the intrinsic properties of materials. However, this enhanced contrast comes at the cost of system complexity, constrained by an inherent trade-off between spatial, spectral, and temporal resolution. To overcome this limitation, we present a self-supervised deep learning-based approach that restores and enhances pixel resolution post-acquisition without requiring external training data beyond the images to be restored. Fine-tuned using metrics aligned with the imaging model, our physics-aware method achieves a 16$\times$ pixel super-resolution enhancement and a 12$\times$ imaging speedup without the need of additional training data for transfer learning. Applied to both synthetic and experimental data from five different sample types, including healthy and diseased tissues, we demonstrate that the model preserves biological integrity, as we did not detect systematic loss of biological features or biologically consequential hallucinations in tested datasets. We also concretely demonstrate the model's ability to reveal disease-associated metabolic changes that would otherwise remain undetectable. Furthermore, we provide physical insights into the model’s inner workings, paving the way for future refinements that could potentially reveal novel high resolution features in an explainable manner. All methods are available as open-source software on GitHub (\url{https://github.com/oycxyd/HyReS}).
\end{abstract}


\section{Introduction}

\begin{figure}[H]
    \centering
    \includegraphics[width=0.6\textwidth]{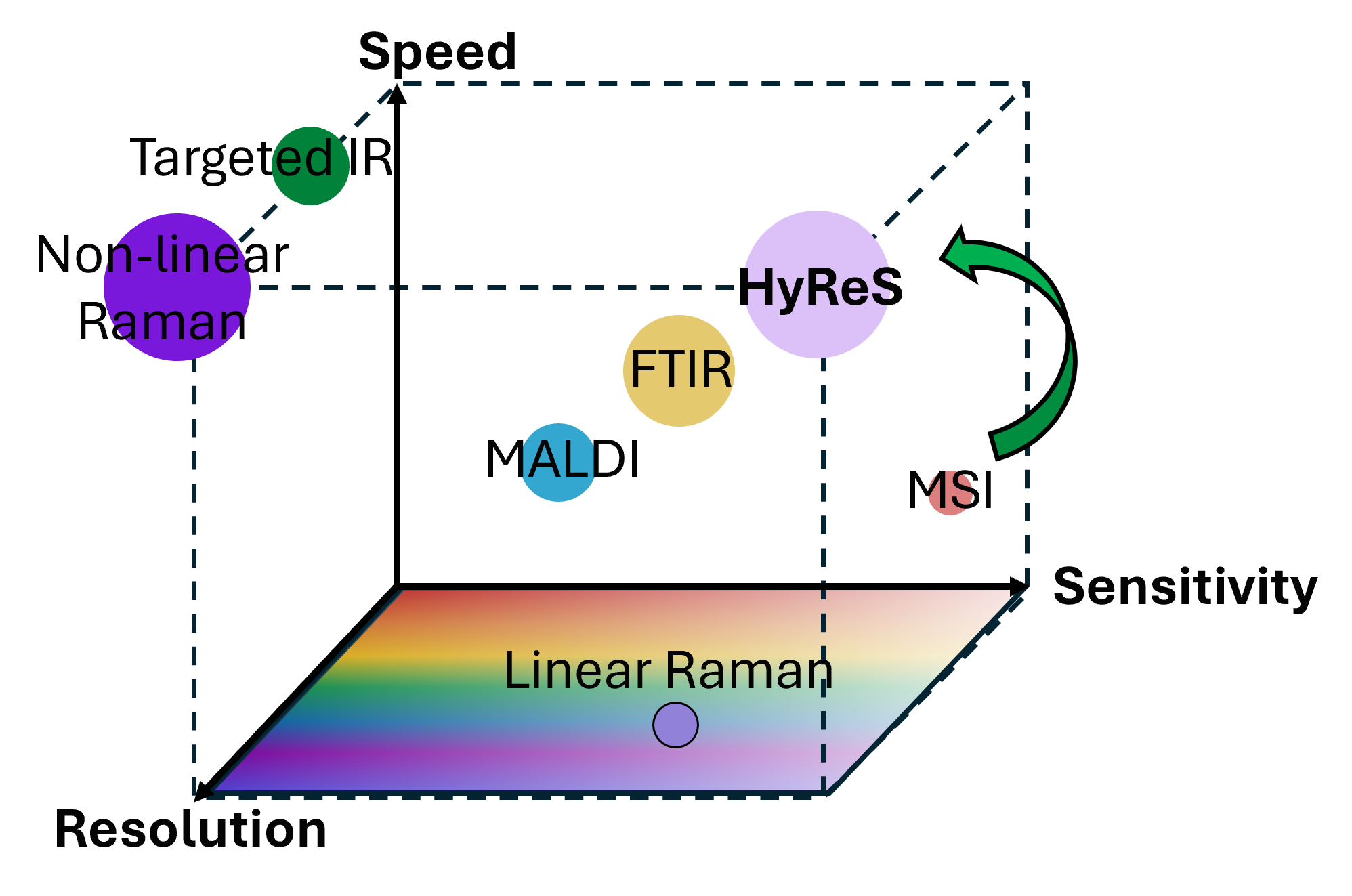}
    \caption{\emph{The 3S space limiting hyperspectral imaging in general and competitor analysis of major approaches granting chemical contrast comparable to that of MSI, with respect to the proposed HyReS scheme. The relative limiting imaging resolution, speed and sensitivity are illustrated by colour-coded bubbles, where darker and larger bubbles indicate higher spatial resolution and faster acquisition respectively, vice versa. The relative transverse locations then denote the chemical coverage in sensitivity. (FTIR: Fourier-Transform Infrared imaging; Targeted IR: Infrared imaging with molecular labels; Linear Raman: spontaneous Raman imaging; Non-linear Raman: stimulated/coherent Anti-Stokes Raman imaging; MALDI: matrix-assisted laser desorption imaging; MSI: all other mass spectrometry imaging techniques.)}}
    \label{fig:3S}
\end{figure}

Hyperspectral imaging (HSI) is a powerful tool for biomedical applications, where the spectral properties of different biological substances can vary significantly. For instance, differences between diseased and healthy tissues can be easily identified, leading to the development and implementation of clinically approved diagnostic and surgical tools~\citep{Fabelo2018, Yoon2022}. In recent years, novel avenues of spectral analysis have been explored, yielding high-contrast images for applications that require biophysical, mechanical, and chemical information~\citep{Singh2020, Kabakova2024, Jones2019a}. These unique mechanisms reveal details that are otherwise invisible with conventional imaging modalities. Despite promising results, the widespread adoption of HSI in biomedicine has been limited. This is primarily due to the well-documented trade-offs between spectral resolution, spatial resolution, and imaging speed (the "3S" triangle), which pose significant challenges, especially in applications requiring fast, real-time operation. Hardware solutions such as spectral scanning and encoding~\citep{Abdo2019} have led to further optimization, but they still cannot overcome the inherent limitations imposed by the 3S triangle (Fig.~\ref{fig:3S}).

Mass Spectrometry Imaging (MSI) is one such HSI technique that has seen significant advances over the last decade because of its unparalleled ability to multiplex thousands of chemical species in a single spectrum and provide their respective spatial mappings. Since its inception in the 1960s, modern MSI can now operate under ambient conditions with minimal or no sample preparation~\citep{Wiseman2006, Simon2025}, making it indispensable in fields such as cancer research, neuroscience, and pharmaceutical research~\citep{Xiao2020}.

Similarly, MSI is constrained by the same triangular limitations. Although MSI generally excels in spectral resolution due to the high sampling rates of common mass spectrometers, its spatial resolution typically falls within the \SIrange[range-phrase={--}]{10}{100}{\micro\metre} range~\citep{Xue2019, Wu2022}, which is significantly lower than that of optical imaging techniques. Achieving higher spatial resolution often comes at the expense of imaging speed. This limitation in spatial resolution may hinder the interpretability of MSI data when combined with currently accepted standards, such as bright-field optical imaging. To address these challenges, existing solutions have often employed image fusion approaches~\citep{VanDePlas2015, Hu2022} to "guide" the localisation of important features. However, these methods are not universally applicable and typically require some level of prior knowledge before fusion can be performed.


To address the intrinsic challenges of MSI, we propose a purely software-based solution in the form of deep learning-facilitated single image super-resolution (DL-SISR). While there is significant interest in DL-SISR within the field of computer vision, particularly for natural scenery images~\citep{Huang2022, Ledig2016}, existing state-of-the-art models typically rely on perceptual metrics-based learning~\citep{Wang2019, Wang2021} that focus on pixelwise differences, without incorporating the underlying physical principles of the imaging process. This makes them less suitable for bioimaging using MSI, where the preservation of spectral and spatial integrity is crucial. Although recent work has implemented DL-SISR algorithms for microscopy~\citep{Weigert2018}, there are two challenges to be highlighted. Firstly, these models primarily enhance perceptual quality rather than achieving true sub-diffraction limit super-resolution~\citep{Qiao2021}. This limitation arises in part from the fundamentally ill-posed nature of the SISR problem, which involves finding a one-to-many mapping from low-resolution (LR) to high-resolution (HR) images. As a related consequence, there is also a high demand for representative training data, which can be challenging to obtain in bioimaging. In contrast, our approach does not need additional training data apart from the data to be restored by synthetically generating naturally paired data in a self-supervised fashion with a physical imaging model. Similarly, we then impose physically meaningful constraints in the Fourier domain to regularise the training process. By incorporating these constraints, our model goes beyond mere perceptual quality improvement, as this physics-based approach helps ensure that the reconstructed images are not only visually accurate but also faithful to the underlying physical properties of the MSI system.

Here, we demonstrate our custom-designed, end-to-end workflow, which facilitates Deep Learning-based Hyperspectral Image Restoration and Super-resolution (HyReS), optimised specifically for MSI. Through various simulated and experimental cases, we show that HyReS functions as an versatile framework capable of both resolution restoration and pixel super-resolution post-acquisition of MSI data, effectively relaxing the practical limitations imposed by current MSI hardware, while maintaining information fidelity for downstream analyses. Crucially, HyReS provides valuable physical insight into whether "true" super-resolution can be achieved, potentially enabling the restoration of novel features that were not discernible in the low-resolution data.

\section{Results}

\subsection{Physics-informed restoration of synthetic data improves image fidelity}\label{section: bc}
We first validate the concept of HyReS by ablation experiments against other state-of-the-art (SOTA) DL-SISR model without incorporating imaging physics (ESRGAN) and traditional method without deep learning (bicubic interpolation).  To-date, robust quality improvement along upsampling for MSI has already been observed, as we~\citep{Xiang2023} and others~\citep{Liao2023} have reported on transfer learning approaches with trained SOTA models~\citep{Wang2019,Weigert2018}. The majority of these models, however, is optimised using perceptual image quality, and then assessed by standard pixelwise metrics such as structural similarity (SSIM) and peak signal-to-noise ratio (PSNR), which only indirectly reflect the image resolution. Here, we propose training our unique Fourier Ring Correlation-based generative adversarial networks (FRCGAN, Section~\ref{sec: methods}) in a self-supervised manner. Unlike other approaches that require well-curated image pairs between different modalities, we generate naturally paired training data by synthetically downsampling with the consideration of detector noise (Section~\ref{sec:evaluation}). We chose a previously reported dataset of human breast biopsies that have corresponding hematoxylin-eosin (H\&E) stained optical images and histopathology assessment\citep{Simon2025}, as a major interest of applying HyReS to MSI is to leverage the speed-up achieved through recovering high-resolution pathology details from low-resolution input taken within a fraction of the time required. The input images were sampled with \SI{50}{\micro\metre} pixel size and their synthetic LR counterparts needed for training are in this case to resemble those sampled with \SI{200}{\micro\metre} pixels. The trained models and interpolation were then used to perform 4-times upsampling with the LR images as input. The results of the top 500 most intense, tissue-specific features are compared. 

\begin{figure}[H]
    \centering
    \includegraphics[width=0.8\textwidth]{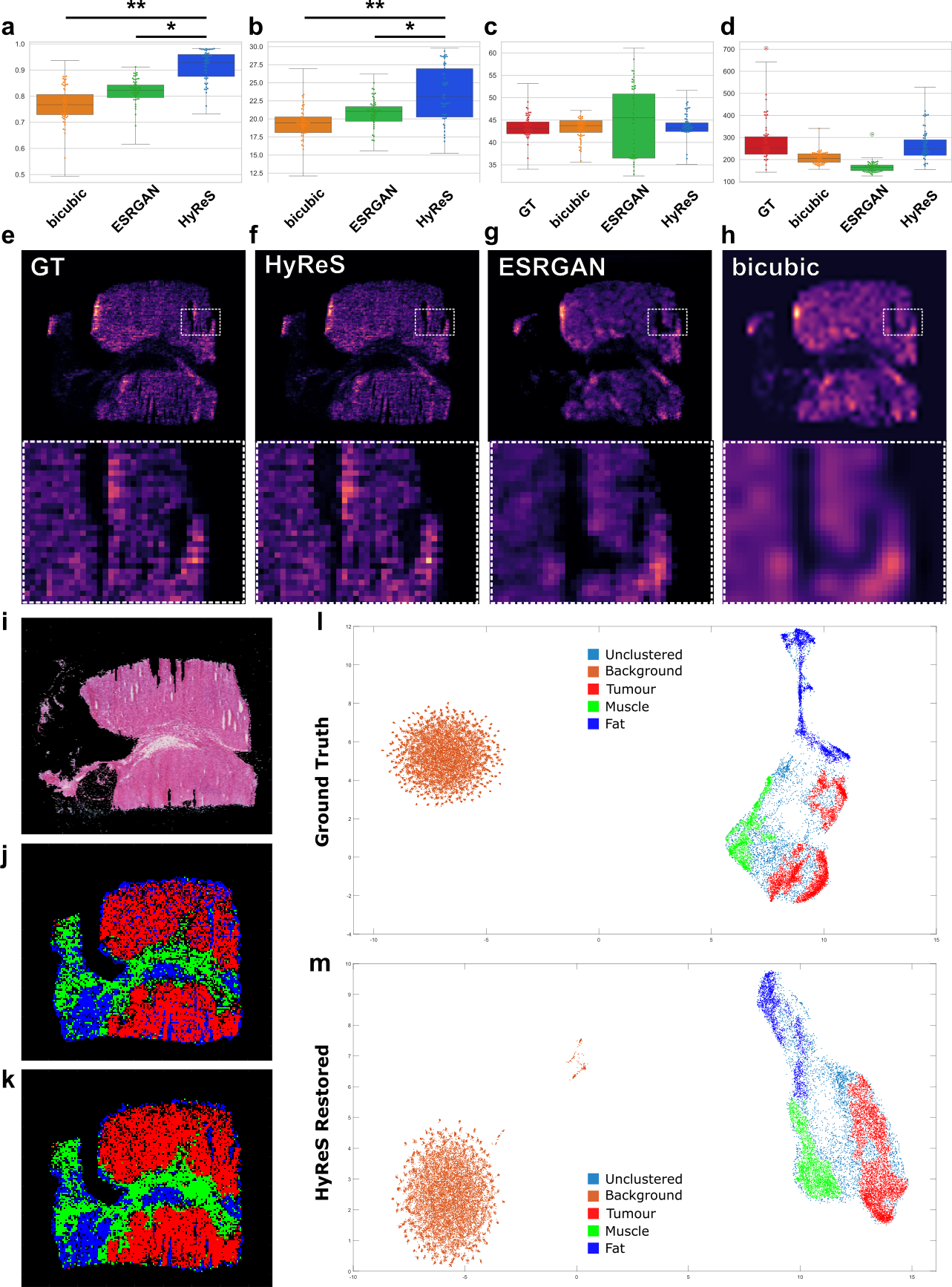}
    \caption{\emph{Restoration of synthetically downsampled breast biopsy images and effect on segmentation. a),b),c),d) summarise the results of the ablation study versus traditional non-DL (bicubic) and SOTA DL models that are physics unaware (ESRGAN), in terms of SSIM, PSNR, perceptual image quality (CRISQUE) and FRC-estimated resolutions respectively. Significantly different values are indicated by *, statistically tested through a two-sided t-test with a corrected p-value threshold of p<0.001. e), f), g), h) present an example of restored images compared to the ground truth, m/z 134.0 (Adenine) was chosen for its relatively high SNR and distinct spatial features. A zoomed-in inset in each case was also included to highlight the performance and behaviour of each method for faithful restoration of biological features, in this case a tissue tear. i), j), k) Comparison of image segmentation results from both the j) GT images and k) those that have been restored by HyReS after downsampling. i) The optical image of an adjacent tissue that has been H\&E-stained is also shown, whose pathological annotations also provided tentative assignment of the segmented regions. Namely, Red = Tumour, Green = Muscle, Blue = Fat. l), m) Comparison of the clustering and hence image segmentation behaviour for both the l) GT and m) HyReS-restored data in a dimensionally reduced 2D domain provided by UMAP. The pathological annotations are illustrated in the same colours as in j,k).}}
    \label{fig:breast}
\end{figure}

The ablation results verify that incorporation of data-specific physical information improves the restorative power, as HyReS outperforms both standard bicubic upsampling and other base DL model in terms of SSIM and PSNR (Figure~\ref{fig:breast}a,b) computed against the ground truth (GT), here referring to the high-resolution datasets acquired at the instrumental limit prior to any synthetic downsampling, serving as the definitive comparative baseline to evaluate the restoration accuracy of the tested models. More interestingly, Figure~\ref{fig:breast}c shows a statistically significant improvement in the image quality of ESRGAN-restored data, compared to both the GT and the corresponding HyReS-restored data. Upon closer inspection, however, the enhancement comes at the cost of hallucination artefact, as demonstrated in Figure~\ref{fig:breast}g. Similarly, while the estimated resolutions are seemingly improved through ESRGAN restoration, this again is due to artificially sharp edges that are not faithful to the original images. Conversely, while there seems to be no statistical difference in quality between the GT and those restored by standard bicubic interpolation, a visible blurriness can be observed in the resultant images due to its fundamentally ill-posed nature (Figure~\ref{fig:breast}h). This meant that the fitting step during resolution estimation (Section ~\ref{sec:evaluation}) generally failed due to a diverging correlation curve, thus a fixed threshold of 0.5 was used to obtain a placeholder number in this case that should not be meaningfully interpreted. Taken together, the results showcase why perceptual quality alone may not be an appropriate metric for restoring resolution and evaluation by SSIM and PSNR do not ensure image fidelity, especially for bioimaging.

As well as the resolution and quality of the restoration, understanding the effect of such approaches on downstream analyses is also crucial for applications such as digital pathology, where objective and trustworthy assessment is key. To simulate such a pathology task, segmentation analysis was performed on both the original GT and HyReS-restored images to facilitate digital `staining' of the tissue into distinct regions of pathological interest. The segmentation was achieved using a mainstream approach that utilises dimension reduction by UMAP~\citep{McInnes2018} and subsequent clustering by HDBSCAN~\citep{EsterM1996}, the parameters were kept identical for both sets of images (Section~\ref{sec:image analysis}). Compared to the H\&E-stained reference (Figure~\ref{fig:breast}a), it can be seen that both the original and restored data give enhanced contrast in segmenting the tumourous (red) and healthy regions on tissue, thanks to the rich spectral information. The latter can be further identified to be either fatty or muscular areas, indicated by the blue and green segments respectively (Figure~\ref{fig:breast}b), agreeing well with the pre-assessment performed on the H\&E data\citep{Simon2025}. To numerically evaluate the effect of HyReS on segmentaiton analysis, we compute the Sørensen-Dice similarity coefficient between the two segmented images, from which an average score of 0.6701 was obtained. While the overall segmentation appears largely similar, the main quantitative difference originates from the muscular and fatty regions. More homogeneous clusters are visibly formed using restored data both in terms of visual perception of the image and their distributions in UMAP space (Figure~\ref{fig:breast}a). Cluster tearing is more obvious in the original data and as such a significant number of data points remain unclustered, which are both respective issues with the UMAP and HDBSCAN algorithms that are well-documented. While the segmentation approach can be optimised parametrically to suit the data, in an unbiased comparison the HyReS-restored data therefore appear more resilient to algorithmic particularity and improves the analysis. Ultimately, the restoration of synthetic data demonstrates the superior fidelity of HyReS over previous models, while the ablation study establishes that physics-aware reconstruction overcomes the limitations of standard perceptual metrics and directly improves the reliability of downstream biological analyses.


\subsection{Restoration of experimental data significantly accelerates imaging speed}
\begin{figure}[h!]
    \centering
    \includegraphics[width=0.8\textwidth]{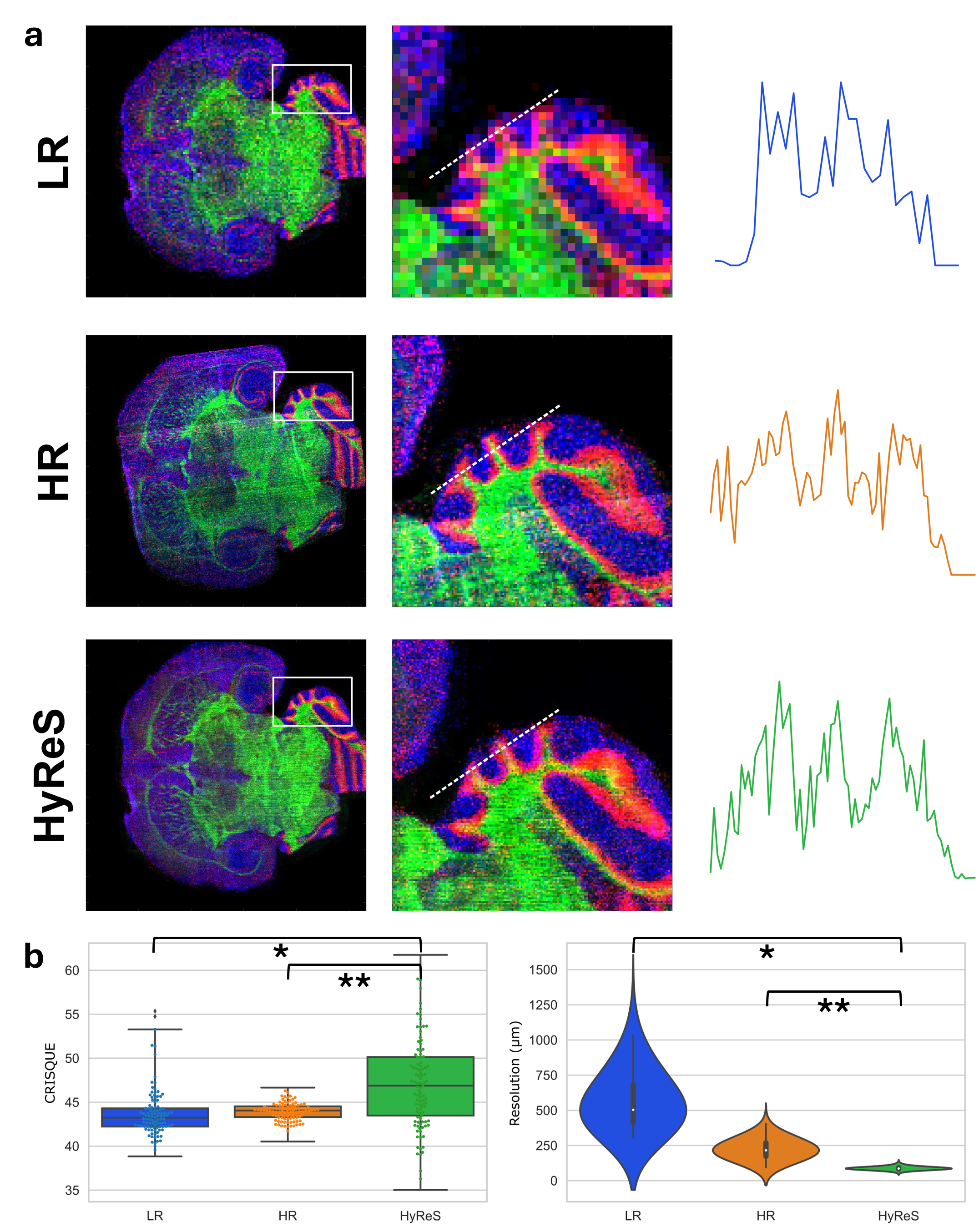}
    \caption{\emph{a) Comparison of RGB ion images (Red:m/z 150.0 Green:m/z 600.5 Blue:m/z 864.6) of mouse brains that have been acquired at low resolution (LR) with \SI{100}{\micro\metre} pixels, at high resolution (HR) with \SI{25}{\micro\metre} pixels and when restored to HR from LR by HyReS. A linescan of the pixel intensities along the cerebellum is also shown in each case to depict the relative SNR. b) Image quality evaluation of bicubically downsampled (LR), ground truth high resolution (HR) and HyReS-upsampled images using CRSIQUE scores and estimated resolutions.}}
    \label{fig:brain_res}
\end{figure}

Encouraged by the restoration fidelity of HyReS on synthetic data, we further investigate if it can be used to restore experimental data with the main motivation of improving imaging speed without compromising resolution. Both HR (\SI{25}{\micro\metre} pixel resolution) and LR images (\SI{100}{\micro\metre} pixel resolution) on adjacent mouse brain sections were acquired to serve as the training and test data respectively. After training a FRCGAN model on the HR data, the LR images were used as input to demonstrate the restoration effect. A summary of the results is visualised in Figure~\ref{fig:brain_res} by means of pseudo-RGB images comprising three representative spectral channels with discernible morphology of the healthy mouse brain. Apart from clearly enhancing the resolution of the LR inputs, the HyReS-restored output images were also observed to have visibly improved SNR and quality when pitched against the HR `ground truth' (Figure~\ref{fig:brain_res}). This reflects directly on the robustness of our approach and its applicability to some MSI-specific properties, for instance the counter-intuitive decrease in SNR and higher possibility of artefact production in HR imaging data due to material removal~\citep{Metodiev2021} and longer imaging time. To quantitatively illustrate this enhancement effect, both the CRISQUE scores and FRC-calculated resolutions are compared for the top 300 highest intensity images in each group. In accordance with the qualitative observation, the image quality increases sequentially going from the LR input to the HyReS output, giving median scores of 43.23, 44.04 and 46.88 respectively. Conversely, the estimated resolutions see a dramatic enhancement from the average of \SI{531.4}{\micro\metre} in the input dataset to \SI{87.2}{\micro\metre} in the restored dataset, with a much tighter distribution, indicating a ubiquitous effect on most input images. In contrast, the GT HR dataset possesses an average resolution of only \SI{212.5}{\micro\metre}, instead of the expectation that it should be four-fold superior to the LR. This can be attributed to the decreased SNR levels and hence lower frequency cut-off points when physically imaging in HR and further strengthens our motivation of a software-based approach to higher resolution.

In terms of imaging speed, the HR data was acquired at a rate of 100 ms/pixel with a total of 207680 pixels, equating to a total run time of around 6 hours. In contrast, at the same sampling rate the LR data only took around 30 minutes and the HyReS restoration an additional 3 minutes. This approach thus represents an order of magnitude effective time save to current hardware-limited MSI without loss of resolution and we believe this to be an exciting new avenue for achieving high-speed imaging in applications where throughput is a major incentive, e.g., digital pathology. These results confirm that HyReS successfully reconstructs accurate spatial architectures within the healthy central nervous system (CNS), establishing a reliable baseline for its subsequent application to complex pathological states.

\subsection{Classification of inflammatory CNS lesions using HyReS-restored images}

After showing the restoration performance of HyReS in the spatial domain of the healthy CNS, we next investigated whether the HyReS-restored images can preserve the information of original images in a pathological context, for which we need a 'ground truth' dataset with known biological implications. We have previously reported on the ability of ambient MSI to metabolically profile in situ the spinal cord of mice with MOG 35-55 experimental autoimmune encephalomyelitis (EAE), a mouse model of multiple sclerosis (MS)~\citep{Peruzzotti-Jametti2024}. MS is a chronic autoimmune disorder of the CNS that affects millions of individuals worldwide. It is characterized by a wide range of neurological symptoms resulting from the attack of the immune system on the CNS, ranging from fatigue, muscle weakness, and vision impairment. While the exact cause of MS is still unclear, the majority of patients eventually develop a progressive accumulation of disability that has been recently linked with changes in tissue metabolism and mitochondrial function~\citep{Peruzzotti-Jametti2021}. Additionally, patients with more progressive forms manifest changes in their neuronal metabolism and mitochondrial function. 

A comparison of spinal cord sections from healthy control (HC) mice and those at the peak of disease (PD, defined as tissue collected three days after the onset of clinical symptoms) demonstrated a clear difference in the intensity distributions of various key metabolites~\citep{Peruzzotti-Jametti2024}. These features were subsequently used to perform predictive modelling with near perfect accuracy scores and serve as an ideal reference to test the biological generalisability of our approach. To evaluate the effect of HyReS on the retention of biological information, the EAE data set was first downsampled and noise-corrupted accordingly, and a pre-trained FRCGAN model was directly applied to the new data to predict their upsampled and denoised counterparts. The same predictive modelling approach was also used to construct classification models based on the downsampled (LR) and restored images respectively using the known features obtained from the ground truth HR images-based model. 

\begin{figure}[h!]
    \centering
    \includegraphics[width=\textwidth]{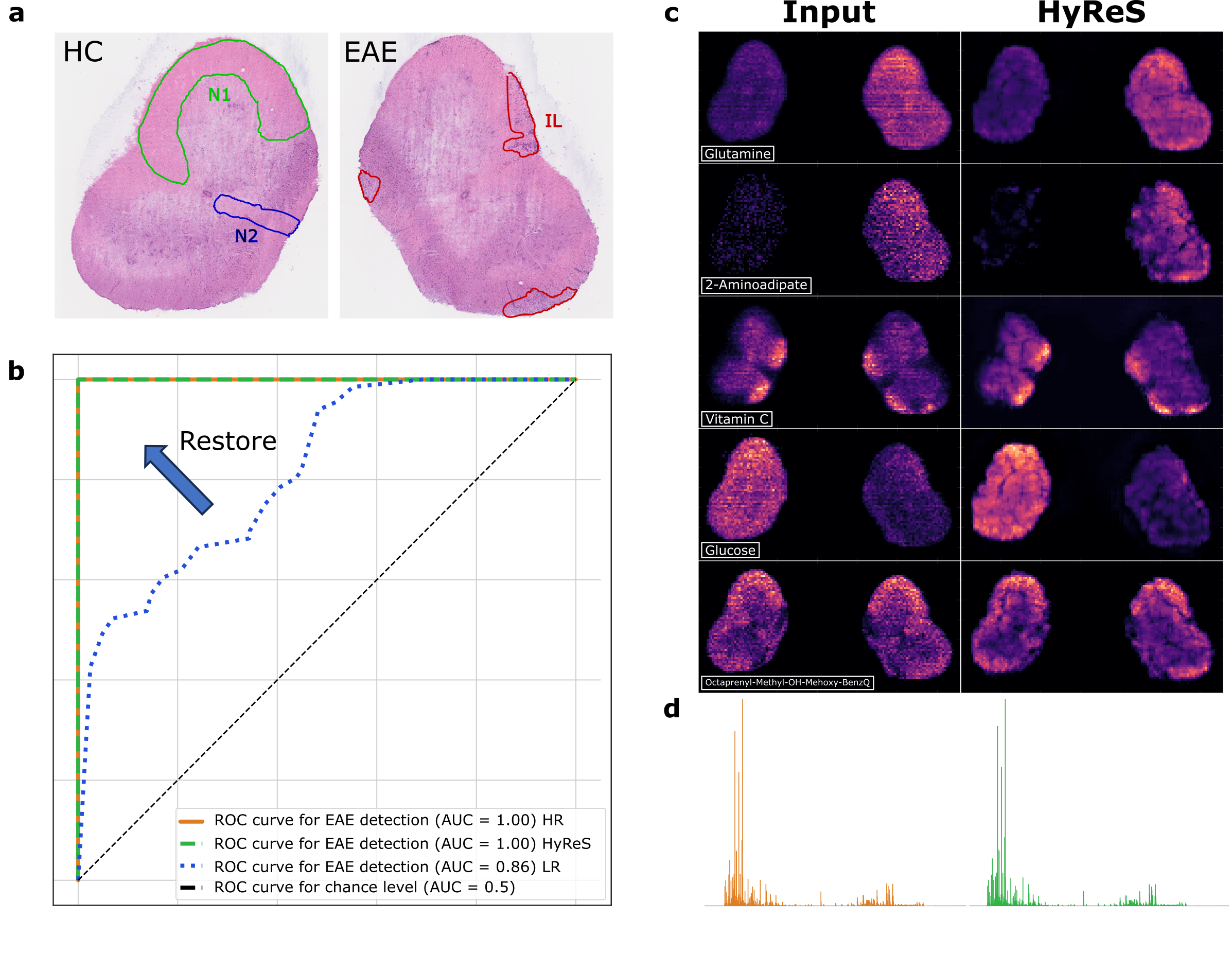}
    \caption{\emph{a) H\&E stained image from an adjacent slide showing regions-of-interest used of predictive modelling, containing areas of AL/DS/Lesion (N1: Normal tissue region 1, N2: Normal tissue region 2, IL: Inflammatory Lesion). b) Predictive modelling results. c) Ion images of key metabolites driving the ground truth classification model before \& after HyReS restoration. d) Mean spectra in the spatial metabolomic data before \& after HyReS restoration.}}
    \label{fig:MSbrains}
\end{figure}

It is well-documented that machine learning models perform optimally when applied to data comparable to those used in training. Although we do not foresee there to be a universal model that can be ubiquitously applied to any new data, we also highlight that the pre-trained FRCGAN still yields high performance in terms of the downstream classification task and does not change the biological conclusion of the study. This is clearly evidenced by the receiver operator characteristics (ROC) curves (Figure~\ref{fig:MSbrains}b), summarising the classification performance on detecting EAE tissue from HC ones. A sharp decrease of $\sim$15\% was induced by artificial downsampling, which was subsequently nearly completely restored by a HyReS model that has not seen this data before. 
Similarly, visual assessment of images of some significant metabolites previously discovered to be driving the classification also show little to no undesirable restoration effect in the results (Figure~\ref{fig:MSbrains}c), while benefiting from some denoising/-artifacting. Some hallucination effect (i.e. addition of new spatial features from the previous training data that are not present in reality) can be potentially identified when comparing with the optical image of an adjacent slide (Figure~\ref{fig:MSbrains}a), though they evidently do not have a major impact on any corresponding image analysis. This is further supported by comparing the spectral characteristics of the original and restored datasets, as their mean spectra were considered to be identical (Figure~\ref{fig:MSbrains}d) according to a Spearman correlation score of 0.9982. The corresponding image quality metrics are collated in the \editx{Supplementary Materials}.

\subsection{Image upsampling for enhanced detection of metabolic phenotypes in a Down syndrome murine model}\label{section:DS}
While the EAE model successfully demonstrates that HyReS preserves macroscopic metabolic signatures in inflammatory lesions, we next sought to determine whether this approach could enhance the detection of more subtle, diffuse metabolic reprogramming. Down's Syndrome (DS) is one of the most common genetic disorders worldwide, which is hallmarked by an excess copy of chromosome 21, hence its alternative name of Trisomy 21. Despite significant progress being made in terms of the genetic understanding that directly leads to observable conditions~\citep{Rachidi2007}, deeper insight is still required to decipher the mechanisms that cause the diverse symptoms of DS, which can be physical, intellectual and behavioural. Recent hypothesis has considered treating DS as a “metabolic disease”~\citep{Dierssen2020}, opening new avenues of metabolic targets that could contribute to understanding pathogenesis, early diagnosis and potential therapeutics. MSI thus provides a unique tool to verify this hypothesis and spatially study the brain metabolic profiles of the Ts65Dn mouse, a trisomic mouse model recapitulating DS-like neuronal, circuitry and behavioural phenotypes~\citep{Rueda2012}. To investigate the metabolic interplay of DS and differences between trisomic mice (Ts65Dn, DS mouse model) and disomic (wildtype) littermates in the first study of its kind, we show that HyReS can be used to elucidate this information from hyperspectral images of mouse brains.

To facilitate a statistically relevant study, a total of $n=20$ mouse brains were sectioned and imaged at a relatively coarse pixel size of $75~\mu m$ in the interest of imaging speed, which will then be restored 4-fold in both dimensions by HyReS post-acquisition. To test the sensitivity of the resultant data to trisomy, similarly sized regions-of-interest (ROIs) from the cerebral cortices in these brain images were taken to train machine learning models. These models were subsequently optimised via feature refinement (Section \ref{sec:data_process}) and then used to predict the aneuploidy-dependent effect of every pixel in the training images ($n=16$) and also independent test images ($n=4$).

\begin{figure}[h!]
    \centering
    \includegraphics[width=\textwidth]{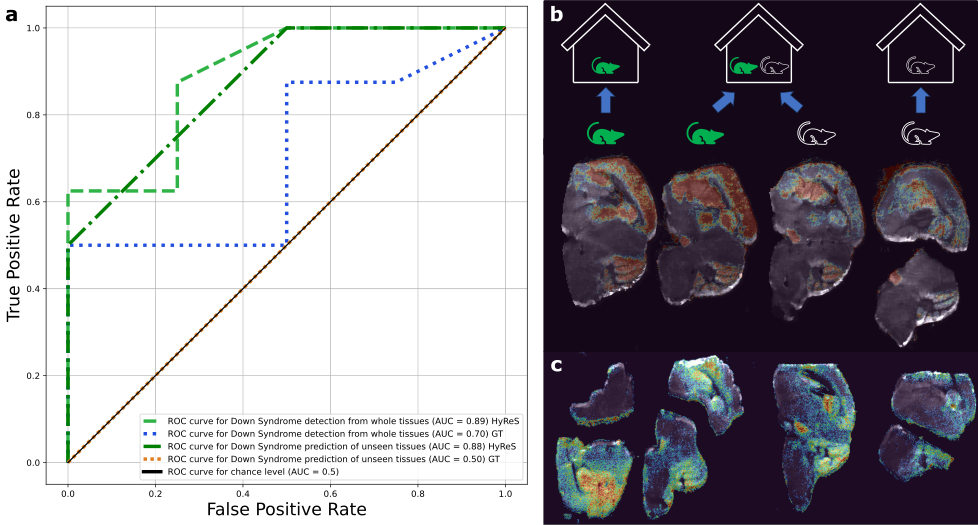}
    \caption{\emph{a) ROC curves summarising the predictive modelling performance of trained models generated from GT and HyReS-enhanced images when applied to validation and independent classification tasks.  b) Representative overlay images showing the spatial distribution and intensity of regions predicted to be trisomic when applying the HyReS-enhanced model on whole brain images used for training. The DS status and caging arrangement of the mice before imaging are also indicated, where green corresponds to a trisomic transgenic mouse, and white a normal mouse. c) Overlay images illustrating trisomic prediction when applying the HyReS-enhanced model on independent whole brain images. The same DS status and caging arrangement from b) also apply.}}
    \label{fig:DSbrains}
\end{figure}

As a result, two binary classification models were trained on the low-resolution `ground truth' (GT) images directly from the experiment and the corresponding high-resolution HyReS-enhanced images respectively, whose performance was evaluated in Table~\ref{tab:clf}. When cross-validated during optimisation, the HyReS-enhanced model notably selects a total of 317 metabolic features deemed to play a significant role in stratifying trisomic and disomic data. This is 180 more than that attained from the GT model, which leads to an apparent 10\% increase in model performance. This performance enhancement is further translated to tissue-based classification when the trained models were applied on whole tissues to predict other pixels that were not seen during training. Namely, the HyReS model robustly achieves an overall accuracy of 82\% (13/16 tissues) while the GT model falls sharply to 63\% (10/16 tissues), suggesting some level of overfitting with the information available. Finally, when given independent classification tasks, the HyReS model retains superior performance at 83\%, only misclassifying one out of four tissues, whereas the GT model is rendered as good as a random coin toss. The quantified assessment of relative performance is also summarised by ROC analysis (Figure~\ref{fig:DSbrains}a).

While the predictive modelling performance serves as a fair comparison, it does not directly reflect the metabolic information of interest, nor biological consideration such as the fact that any disease-associated abnormalities are not necessarily found in the entire brain. In which case, it is more insightful to interpret the spatial and spectral features involved in classification. To visualise the prediction spatially, the probabilities of pixels being predicted to be trisomic were mapped and overlayed with an image from a channel (m/z 885.54 PI(38:4)) is the MSI dataset that represents the anatomy well. Figure~\ref{fig:DSbrains}b\&c demonstrate such overlays of HyReS-enhanced images where the brains from trisomic mice can be clearly distinguished from those from disomic mice due to higher prediction probabilities both in value and in density. Furthermore, despite training the model on pixels from the cerebral cortex alone, the prediction shows localised `hot spots' that are anatomically specific across the brain, highlighting e.g., the dorsal striatum, basal forebrain, and cerebellum. This is in line with expectation as these components play roles in both physical and cognitive functions, which are directly affected by trisomy 21~\citep{Dierssen2020}. More interestingly, the housing arrangement of the experimental mice may also introduce a social element that is further revealed by HyReS, as some trisomic and disomic mice were caged together for the purpose of a behavioural study before the imaging experiment. This arrangement was also reflected in the prediction results, as disomic mice that lived alongside the trisomic ones appeared to be more prone to miclassification. In fact, all of the misclassified cases arising from the HyReS models were from co-caged disomic mice, predicted to have a relatively higher concentration of `trisomic pixels'. Specifically, there seemed to be a high localisation in the dorsal striatum, which has also been linked to habit formation~\citep{Goodman2018}. 

\begin{figure}[h!]
    \centering
    \includegraphics[width=\textwidth]{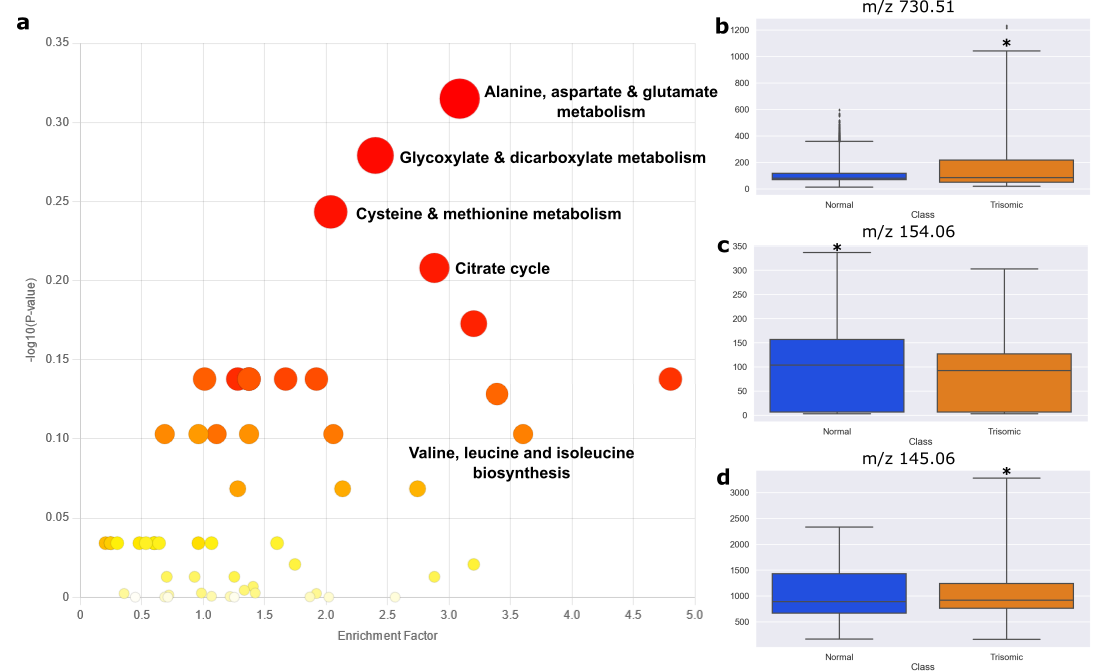}
    \caption{\emph{a) Metabolic pathway enrichment analysis using the KEGG online database. The logged p-values from the statistical analysis for each metabolic network is indicated on the y-axis, while their corresponding enrichment factor is indicated by the x-axis. b)-d) Representative metabolic features that are significantly up-regulated, down-regulated and statistically unchanged respectively between DS and normal mice. m/z 730.51 is tentatively identified as a ceramide which is invovled in inflammatory responses in diseases comorbid of DS~\citep{Worley2023}; m/z 154.06 is tentatively annotated to be histidine, which has been cited to be reduced significantly in DS human~\citep{Caracausi2018, Dierssen2020}; m/z 145/06 is tentatively identified as glutamine, which has been observed to be unperturbed in terms of bioenergetic pathways of DS~\citep{Pecze2020}.}}
    \label{fig:DSfeatures}
\end{figure}

Spectrally, HyReS provides a list of significant features that facilitates further interpretation which may provide biological insight. Even without the exact identification of chemical species, functional interpretation is already possible through algorithms such as mummichog~\citep{Li2013}, which statistically predicts the activity of metabolic pathways and networks in play. Figure \ref{fig:DSfeatures} thus shows the result of such functional analysis which highlights intensity variations that may implicate e.g., Lysine degradation, Valine, Leucine, and Isoleucine biosynthesis and degradation, Glyoxylate and Dicarboxylate metabolism, Tyrosine metabolism, Phenylalanine metabolism and the Citrate cycle. Despite the spuriously high p-values that are due to a limited degree of identification (217 hits from 62 masses), these findings are consistent with previous work. Namely, a recent study by Cai et al.~\citep{Cai2023} revealed very similar pathways in mice with Down's Syndrome, which also affected their social behaviour. In features that have been identified to be significantly up- or down-regulated, potentially identifiable amino acid species are also in line with the trends observed previously in human subjects~\citep{Caracausi2018}, while heavier lipid species reflect trends that have been linked to comorbidities of DS~\citep{Worley2023}.

\begin{table}[h!]
\centering
\caption{\emph{Performance comparison of Down Syndrome detection with models trained on regions from the cerebral cortices of the original and HyReS-enhanced images.}}
\label{tab:clf}
\begin{tabular}{lllll}
\hline
         & \textbf{Sensitivity}    & \textbf{Specificity} & \textbf{Balanced Accuracy} & \textbf{AUC} \\ \hline
\textbf{GT} (10-fold CV, 127 features)    & 0.80 & 0.80    & 0.80 & 0.80       \\
\textbf{HyReS} (10-fold CV, 317 features)    & 0.90 & 0.90    & 0.90 & 0.90     \\ 
\textbf{GT} (whole tissue prediction)    & 0.60 & 0.67    & 0.63 & 0.70      \\
\textbf{HyReS} (whole tissue prediction)    & 0.85 & 0.78    & 0.82 & 0.89       \\
\textbf{GT} (independent prediction, )    & 0.50 & 0.50   & 0.50 &  0.50  \\
\textbf{HyReS} (independent prediction)    & 1.00 & 0.67 & 0.83 & 0.88       
\\ \hline
\end{tabular}%
\end{table}

In summary, HyReS enhances the spatial and spectral sensitivity for downstream analyses, achieving better results in binary classification than that was obtainable in the original dataset. The spectral improvement can be directly linked to the denoising effect of HyReS which in turn resolves the more subtle intensity differences. In terms of the spatial resolution, HyReS appears to offer more than just denoised, upsampled training data. The restored images have  higher spatial frequency cut-offs and higher visual quality scores overall (\editx{Supplementary}),
which seems to suggest an enrichment, rather than linear increase in spatial information. While the spatial features in this case are of an anatomical scale, it is foreseeable that HyReS may have a major impact on analyses where the resolution of features close to the instrumental limit is vital.

\subsection{Pixel super-resolution enables multi-omic integration across scales}\label{section:ffpe}
On top of spatial resolution, the main attraction of using MSI for biomedical applications remains the additional spectral contrast. To gain insight into the resolution enhancement enabled by HyReS relative to other concurrent understanding in the field, we apply FRCGAN to clinical, formalin-fixed paraffin-embedded (FFPE) samples in the form of tissue microarrays (TMA), which has been collected as a part of the larger cohort TransSCOT -- a study on thousands of colorectal cancer patients with comprehensive clinicopathological metadata~\citep{Iveson2018}. These FFPE TMAs are especially interesting here as the multi-modal spatial analysis of individual cores is expected to expose patient-level heterogeneity of the disease by bridging gaps between multiple sources of complementary information, including genomics, transcriptomics and metabonomics. While ambient MSI can be used to excavate metabolic information, it has traditionally been considered incompatible with FFPE-preserved samples due to a degradation in the signal intensity. This coupled with the order of magnitude mismatch in spatial resolution with corresponding data from techniques such as imaging mass cytometry (IMC) \& immunohistochemistry (IHC) are an ideal application for the HyReS approach. 

\begin{figure}[h!]
    \centering
    \includegraphics[width=\textwidth]{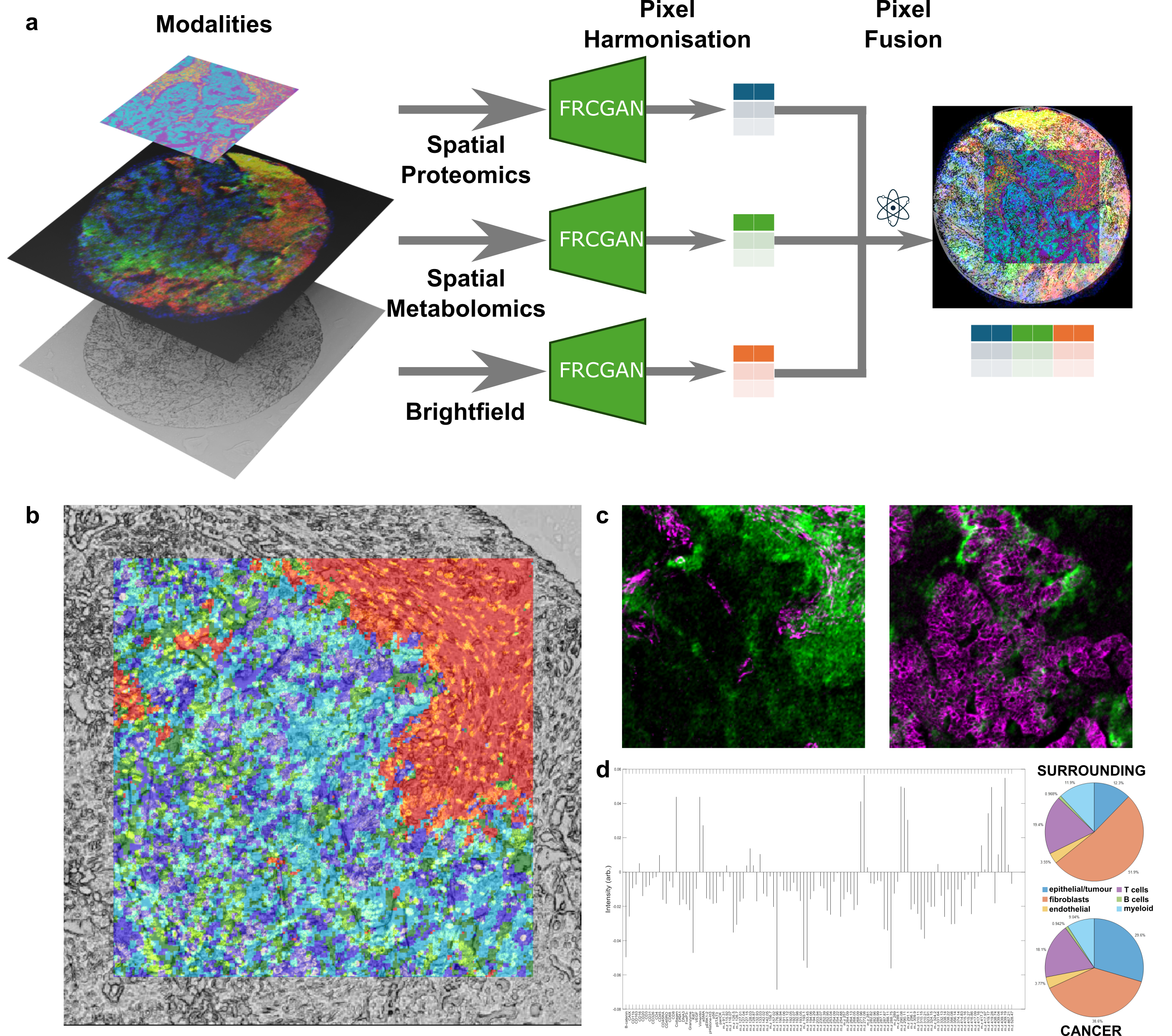}
    \caption{\emph{Application of HyReS to single-cell level spatial omics. a) Pixel-wise image fusion strategy employing HyReS to facilitate pixel resolution harmonisation. The abundance/intensity per pixel is concatenated after multi-modal fusion to create multi-modal contrast and producing a single data matrix for further analysis. b) Segmentation analysis performed on the resultant multi-modal matrix to digitally enhance the grayscale contrast of the corresponding brightfield image. Four distinct clusters (red, green, blue, cyan) are highlighted spatially due to their multi-omic profiles, while a DNA-specific marker (yellow) was also used to co-visualise the individual nuclei for cell counting. c) HyReS-restored images of metabolites which shows high spatial correlation with certain IMC channels discovered by spatial clustering, (left) the ion m.z 290.08 overlayed with a fibroblast marker (aSMA). (right) the ion m.z. 192.98 overlayed with an epithelial/tumour cell marker ($\beta$--catenin) d) Multi-omic differential profile calculated from the average expressions of the cancer (cyan, green, blue) and surrounding (red) regions in b). The total number of cells in each region is estimated by nuclei counting and classified into tumour/epithelial, fibroblast, endothelial, T, B and myeloid(-like) cells. The relative cellular compositions in percentages are also shown alongside for easy comparison.}}
    \label{fig:ffpe}
\end{figure}

Using a high-resolution imaging setup we have developed in-house with a picosecond laser~\citep{Battle2023}, a single TMA core has been imaged by LD-REIMS~\citep{Simon2025} with a pixel size of \SI{10}{\micro\metre} and subsequently upsampled by FRCGAN four times on both axes (Figure~\ref{fig:ffpe}). The network was trained and optimised using our standard approach (Section \ref{sec:data_process}). Owing to the 4$\times$ upsampling, a theoretical pixel resolution of \SI{2.5}{\micro\metre} brings the MSI image to be much more on par with the optical equivalent, which has a pixel resolution of $\sim$\SI{1}{\micro\metre}. In tandem, IMC was performed with a panel of 35 metallic labels on the same patient core form an independent section, a square ROI of \SI{500}{\micro\metre\squared} was sampled also with a \SI{1}{\micro\metre} pixel size. By harmonising pixel sizes across modalities to\SI{2.5}{\micro\metre}, direct multi-modal fusion was made possible. The integrated data were subsequently segmented using 116 multi-omic targets deemed to be highly co-localised (\ref{sec:data_process}) by an unsupervised approach into 4 endmembers for visualisation and qualitative multi-omic interpretation. 

The integrated images allow direct visual inspection of the overlap between modalities across scales (Figure~\ref{fig:ffpe}a). As we do not expect cell-to-cell overlap between tissues, we instead highlight two distinct regions discovered through the multi-omic contrast. Namely, as the IMC panel encompasses a broad coverage of tumour–immune microenvironment and signalling, the red region in Figure~\ref{fig:ffpe}b correlates well with the presence of myeloid, T cells and subset, as well as fibroblasts. Interestingly,  the most co-localised metabolite is m.z 290.08 (Figure~\ref{fig:ffpe}c), which is tentatively identified as N-Acetylneuraminic acid, boosted in confidence given the well-described role of sialic acid in immune cell activation. In contrast, the cyan, blue and green segments combined overlaps with markers for tumour/ epithelial cells, as well as stromal and extracellular components in the microenvironment. In this case, a highly correlated metabolite is m.z 192.98 (Figure~\ref{fig:ffpe}c), believed to be arising from the [M+Cl] adduct of the compound $C_6H_6O_5$. 

In addition to discerning the overall heterogeneity in the tumour microenvironment, the multi-modal information now allows quantification at the single-cell level. Whilst exact registration of individual cells is not possible due to misalignment in orientation \& minor differences in structural details between adjacent sections, estimation of cell numbers in each of the identified phenotypic region is possible. For example, by highlighting the nuclei using the Ir193 channel (yellow in Figure~\ref{fig:ffpe}b), rough number and types of cells in the tumour and surrounding microenvironment can be approximated using specific IMC markers. A difference profile between these regions can then be computed (Figure~\ref{fig:ffpe}d), which directly visualises the up- or down-regulation of the markers covered in the integrated proteo-metabolome. When linked to the relative abundance of certain cell types, we observe that the cancer region see fewer T, B, and myeloid-like cells and more tumour‑like epithelium. This often reflects an immune‑excluded phenotype, where lymphocytes are pushed into the surrounding stroma or margin rather than infiltrating the tumour core. Further supporting this hypothesis, fibroblasts also dominate the surrounding stroma, which could imply extracellular matrix remodelling which causes the impairment in immune cell penetration and hence worse prognosis\citep{andac-aktas_immunological_2025}. The biochemical mechanism of such metabolic rewiring can also now be studied simultaneously from the integrated multi-modal information from just a single core. Scaling this up to the cohort and even population level from widely available FFPE TMA samples for high-throughput molecular screening would be of direct interest and in line with the motivation of the TransSCOT study. This demonstrates to our knowledge the first example of such spatial information extraction at a cellular level and thus offers the direct route to single-cell omics in a biological context. We thus foresee HyReS a practical route towards spatial multi‑omic integration at near single‑cell resolution, synergistically enhanced in the near future by data acquisition methods that would provide multi-source information from the same slide\citep{potthoff_spatial_2025, nunes_integration_2024}, applicable to numerous systems biology approaches.

\subsection{Assessment of resolution enhancement with physical explainability}
\begin{figure}[h!]
    \centering
    \includegraphics[width=\textwidth]{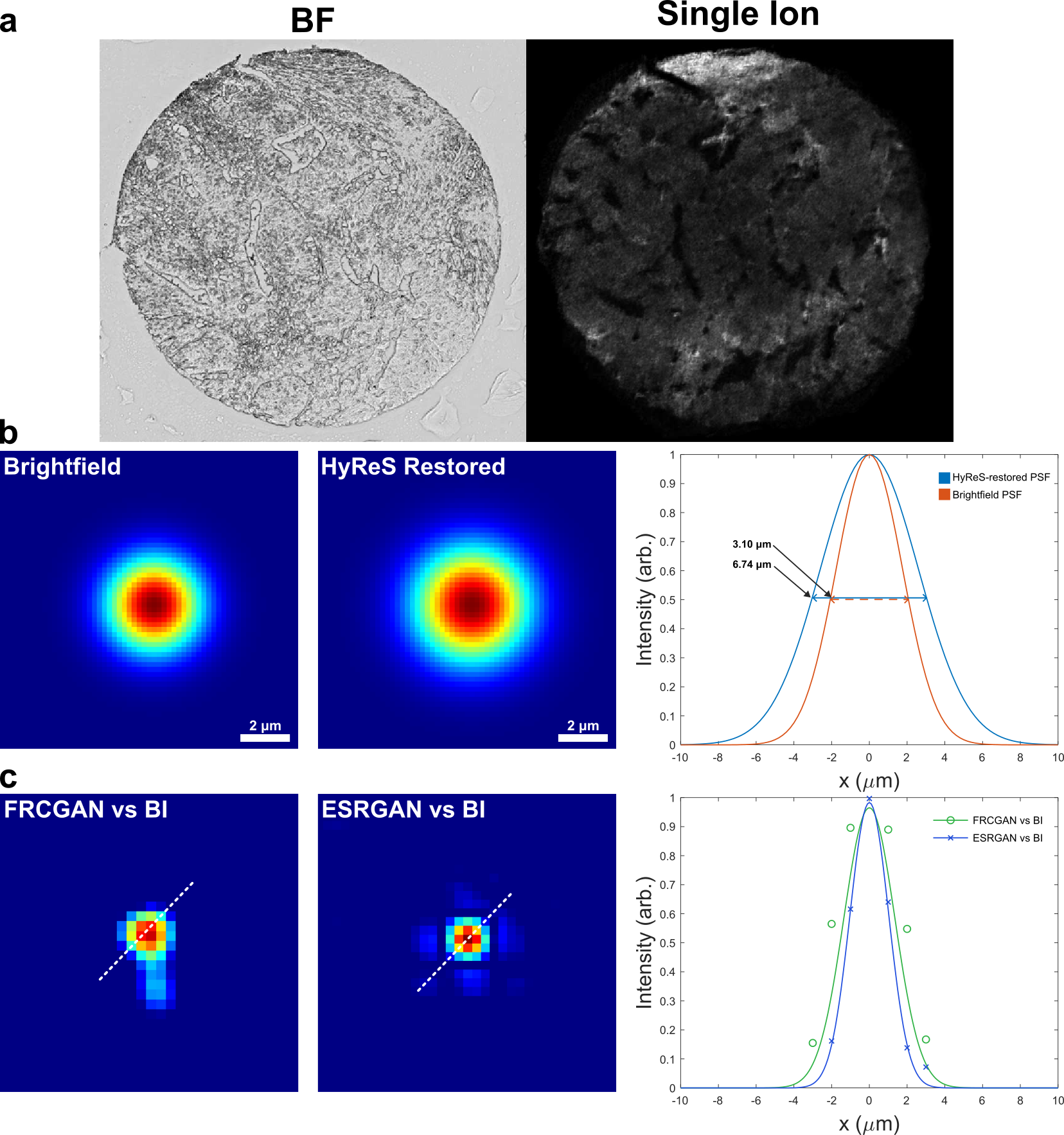}
    \caption{\emph{Evaluation of the physical effect of applying HyReS. a) Comparison of super-resolved FFPE core image (m.z 128.03) with the corresponding brightfield optical image. b) the 2D PSFs estimated from the HyReS-embedded FRC measure for images shown in a) and Gaussian line fits to extract the FWHM and hence resolution of the images. c) the difference PSFs obtained from the the same image upsampled by FRCGAN/ESRGAN with its bicubically interpolated counterpart. 2D fits Gaussin fits display a larger FWHM for FRCGAN, indicating a stronger deblurring effect.}}
    \label{fig:psf}
\end{figure}

To interpret the strong restorative ability of FRCGAN, we further investigate the physical effect of the model thanks to the built-in single-image resolution estimation, and subsequently evaluate of the possibility of `true' super-resolution, i.e., recovery of novel objects within the image, using only HyReS. Compared to standard microscopy, some HSI modalities are limited in spatial resolution by physical phenomena that give rise to their contrast mechanism, such as the phonon interaction length~\citep{Passeri2023}. In MSI, this is largely instrument-dependent and may even be user-dependent, with traditional methods e.g., MALDI affected by the quality of the matrix. To this end, significant effort has gone towards the development of ambient MSI modalities which require minimal or no sample preparation, making them favourable for direct analysis from soft, biomedical samples. The current technology is however cited to be still lacking in spatial resolution~\citep{Xiao2020}, thus struggles to provide cellular (\SI{}{\micro\metre}) level detail that is often essential for life scientists and clinicians. While we have previously concluded that SOTA SISR models, such as ESRGAN are unlikely to produce any enhancement beyond the instrumental limits~\citep{Xiang2023}, we revisit the same challenge with the well-trained FRCGAN. Using the model-inherent FRC measure to independently evaluate the effective resolution before and after performing restoration on a representative FFPE TMA image from Section \ref{section:ffpe} (m.z 128.03), values of \SI{32.92}{\micro\metre} and \SI{6.74}{\micro\metre} were yielded respectively, meaning an approximately four-fold enhancement was indeed observed. These numbers are effectively auto-computed by FRCGAN during training, which also allows retroactive estimation and visualisation of a pseudo-PSF that is physically meaningful. For reference, we show the FRC-estimated PSF of the corresponding brightfield image alongside that of the HyReS restored hyperspectral image(Figure~\ref{fig:psf}b). The former, sampled with \SI{1}{\micro\metre} pixels, can be seen to attain an effective resolution of \SI{3.10}{\micro\metre}, suggesting that the restoration result is approaching, if not exceeding the pixel-limited resolution of the physical imaging model. Due to the nature of our FRC calculation strategy, however, some uncertainty especially when comparing single images are expected, due to for example, under-sampling (Section~\ref{sec: methods}). 

To remove any method-dependent bias and to visualise the inner-working of the Fourier imaging model underlying FRCGAN, we compute the difference PSF from the results (Section~\ref{sec: methods}). First we compare the difference PSF between the FRCGAN-restored image and the original image before upsampling, with the latter resized to the same dimension by simple bicubic interpolation. Interpolation of this kind is known to induce additional blur~\citep{Liu2013} and unlikely to provide any gain in resolving power, as such a largely Gaussian difference PSF is obtained (Figure~\ref{fig:psf}b), agreeing with the Fourier model prediction. Similarly, to benchmark our method against the closely related SOTA -- ESRGAN, we perform the same calculation between resultant images generated by the two models (Figure~\ref{fig:psf}c).  While an algebraic ambiguity exists in the sign of the difference calculation, i.e., which PSF is wider, we elucidate this by rationalising a positive difference PSF between both the ESRGAN-restored image and FRCGAN-restored image against the interpolated `ground truth'. By inspection, a discernibly Gaussian distribution is still observed in the resultant ESRGAN PSF but with narrower width, indicating much more comparable PSFs in the reconstructed images, indicating inferior deblurring performance. To quantify this, 2D Gaussian fits were carried out on the PSFs compared assuming a rotationally symmetric shape for simplicity (Figure~\ref{fig:psf}c). As a result, it can be shown that FRCGAN outperforms base ESRGAN by a factor of $1.31$ in terms of fitted PSF width. Furthermore, the recovery of the predicted Gaussian functions serves as direct evidence that FRCGAN and to some extent ESRGAN models are performing deconvolution procedures that are routinely used in commercial microscopy to super-resolve images~\citep{Borlinghaus2016}. Rather than purely enlarging the images, the models have also acquired the ability to remove the effect of a blurry PSF according to the training data, which reiterates the importance of machine learning physically meaningful (PSF-relevant) trends in applying the HyReS approach.

\section{Discussion}
We have demonstrated that HyReS networks trained using the FRCGAN framework can be directly applied to a range of biomedical applications without any prior information nor need of additional training data e.g., well-curated optical images of the same tissue type. The training strategy makes use of widely applicable inter-data correlations within hyperspetral datasets, which can be readily used to train high-performance models by introducing the understanding of the imaging and noise model without the need of large quantity and variety in training data generation.

We show that HyReS can be implemented to address the limitations imposed by the 3S conditions that still restrict MSI, and HSI in general. Both synthetic and experimental images across MSI modalities and tissue types can be restored from its 4-times downsampled input with no obvious degradation to image resolution or quality. In contrast to other DL-SISR models available, we show that by moving away from an over-emphasis on perceptual image quality, the FRCGAN architecture preserves both spatial and spectral information, which can be further verified by downstream analyses that ultimately incentivise the use of MSI, such as segmentation.  

The sequential application of HyReS across these diverse CNS models confirms its biological validity. In the healthy CNS, the model accurately mapped basal metabolic distributions without introducing artificial features. In the inflamed CNS, it preserved the critical metabolic shifts associated with neuroinflammation in EAE. Furthermore, in the genetic model of Down syndrome, the enhanced resolution provided by HyReS uncovered subtle metabolic vulnerabilities, highlighting its utility in dissecting complex neurological diseases across different scales of pathology. 

We thus foresee the use of HyReS in combination of concurrent mass spectrometry imaging systems to have a major impact for high-throughput applications, such as histopathology. By pre-training networks using images from a higher pixel resolution, HyReS is able to predict and upsample new unseen images taken with larger sampling pixels. This results in an increase in the imaging speed of at least 16 times ($\times$4 on both axes), allowing point-scanning MSI modalities such as DESI to produce whole tissue images in minutes rather than hours, without compromising spatial resolution. This approach requires, however, a model trained on an experimentally generated high-resolution dataset, so it does not establish performance without initial additional training data. Even higher upsampling ratios can of course be explored with HyReS, but given the ill-posed nature of this restoration problem~\citep{Tipping2003}, deeper understanding of the imaging model will likely be required as prior information, both in terms of training data generation and the imaging formation process that needs to be incorporated in training.

As well as spatial resolution restoration, we have shown HyReS to have a positive impact in general on downstream image analysis without significantly altering the information content. Apart from the visually apparent denoising effect that has been proven to be generalisable to datasets of diverse origins~\citep{Xiang2023}, the single image super-resolution process also grants increased data points (pixels) for machine learning tasks while restoring features that are otherwise corrupted by under-sampling, noise and artifact. This has been highlighted by the example of enhanced rate of detection in the metabolic difference in normal and trisomic mouse brains, where the HyReS-enhanced regression model outperforms the original using classification features that are still interpretable. Spectrally, HyReS provides a list of putative features; orthogonal MS/MS and quantitative assays are still required to confirm identities and effect sizes. To separate the effect of HyReS and the complex biological processes of interest in such problems, we have also demonstrated the strong generalisabiltiy and robustness of trained HyReS models by directly applying them on unseen data from a study of known conclusions. The preservation of the metabolic profiles in the restored mouse brains and the general observation of a strong contrast in the late-stage multiple sclerosis model due to a few identified molecules speak volume to the wide utility of HyReS in future studies of this kind. While we do not recommend using the same HyReS model for all tasks, we note that the minimum requirement of HyReS in training data and its self-supervised nature largely reduces the computational complexity of fine tuning new models, when the data changes drastically e.g. in new tissue types. Online training~\citep{liu_survey_2021} using a mixture of new and existing data may also be an interesting approach to accumulate restoration power and to minimise hallucination, especially with fewer samples and data points.

Finally, echoing some existing concerns on the matter~\citep{Xiang2023,Qiao2021}, whether HyReS is able to restore novel features i.e., achieving true super-resolution is less certain. To this end, we have investigated the relative effect on the image PSF induced by HyReS in accordance with our built-in Fourier imaging model. It is clear that an embedded deconvolution effect is taking place and the quantifiable deblurring achieved in turn amplifies the high spatial frequency content that is already in the image. Interestingly, we note that this subtlety is generally omitted or misunderstood in existing methods, as others have attempted to include our imaging model by convolving already blurry images with an artificial PSF that does not reflect the instrumental behaviour\citep{Liao2023}. Akin to confocal microscopy, this may provide more resolving power even though the hardware remains unchanged provided that some signal collection selectivity is in place. In terms of MSI, we have so far omitted the collection path in our proposed imaging model and no such confocal-like models have been formulated to our knowledge, which motivates future work. By measuring PSF information experimentally, sufficient prior knowledge may be obtained to transform HyReS into a true super-resolution HSI tool to go beyond the hardware limitations. Going forward, we envision the combination of HyReS with the development of its corresponding imaging hardware to provide the platform for multimodal studies across scales in the same vein as the FFPE core example given here. Similarly, HyReS can be easily applied to other modes of HSI, as we already demonstrate comparable results (\editx{Supplementary}) in e.g. Brillouin microscopy~\citep{Kabakova2024} with minimal modification and hence unveils new possibilities that simultaneously demand imaging throughput \& resolution.

\section{Methods}\label{sec: methods}
In this work we aim to learn the mapping from the LR images to their respective HR counterparts while achieving two main restorative effect: denoising and pixel super-resolution. In this section, we briefly introduce three essential components of our study: (1) the imaging experimental methods, (2) the SISR models and (3) the evaluation methods for images and trained models. 


\subsection{Ambient Mass Spectrometry Imaging}
\subsubsection{Laser-Desorption Rapid Evaporation Ionisation Mass Spectrometry (LD-REIMS) imaging}\label{subsec:reims}
\textbf{Human breast biopsies for synthetic image restoration}
For LD-REIMS imaging, all experiments utilised a commercially available Opolette HE2731 Optical Parametric Oscillator (Opotek, Carlsbad, USA) running at 20 Hz and tuned to \SI{2.94}{\micro\metre} for mass spectrometry imaging. A distal optical system, assembled from optomechanical components (Thorlabs Inc.) and a CaF2 lens (LA5315-E) focused the beam onto the sample plane with a diammeter of $\sim 60 \mu m$. Imaging experiments were conducted using a modified two-dimensional stage setup (Prosolia), connected to a Xevo G2-XS QTOF mass spectrometer (Waters Corporation, Milford, MA, USA) controlled by MassLynx 4.1 software (Waters Corporation, Milford, MA, USA). The mass spectrometer featured a prototype REIMS source~\citep{Jones2019c}. All mass spectrometric analysis was conducted in the negative ion mode within a spectral range of 50-1200 m/z.
To demonstrate the concept of image restoration by HyReS, human breast biopsies were imaged using this setup with a pixel size of $50 \mu m \times 50 \mu m$. The resultant images were then downsampled synthetically before restoration during preprocessing (Section \ref{sec:data_process}).

\textbf{Adjacent mouse brain sections for experimental image restoration}
To experimentally validate HyReS, mouse brain sections were imaged using the same in-house setup. An aspheric lens (C028TME-E, Thorlabs Inc.) was utilised instead of the CaF2 singlet to correct for aberrations. The new beam diameter was measured to be approximately \SI{30}{\micro\metre} at the sample plane. 
To create the LR and HR pair, adjacent brain slices that were morphologically similar were independently imaged with pixel sizes of  $100 \mu m \times 100 \mu m$ and  $25 \mu m \times 25\mu m$ respectively.

\textbf{High-resolution LD-REIMS imaging of FFPE TMA core}
For high-resolution, a prototype optical parametric amplifier (OPA) system operating in the picosecond regime~\citep{Battle2023} was used in place of the OPO from previous LD-REIMS experiments while all other components were identical. Briefly,the OPA system consists of a high power pump laser and a nonlinear crystal to achieve amplification through parametric interaction. A pulsed microchip laser operating at \SI{1.06}{\micro\metre} with a repetition rate of 500 kHz is amplified by an ytterbium (Yb)-doped fiber amplifier (YDFA). The output from the YDFA is used as the pump laser, providing a pulse energy of 6 $\mu J$ for the amplification process. A seed signal at a wavelength of \SI{1.67}{\micro\metre} is introduced into the nonlinear crystal, which is magnesium (Mg)-doped Periodically Poled Lithium Niobate (PPLN) crystal. The crystal is designed with a poling period of \SI{31}{\micro\metre} to phase match the pump and seed wavelengths, enabling efficient energy transfer from the pump to the seed and generating an amplified output signal and an idler at the wavelength of \SI{2.94}{\micro\metre}. The pump and the seed beams are focused into the PPLN crystal using an N-BK7 lens with f = 125 mm. The PPLN crystal is mounted on a temperature-controlled oven and held at a fixed temperature of 164 ℃ to maintain optimal phase matching conditions. The output beams are then collimated by a calcium fluoride (CaF2) lens with f = 200 mm for further use in the REIMS imaging. The idler is centered at \SI{2.94}{\micro\metre} with a pulse duration of ~100 ps, and a pulse energy of 400 nJ is delivered to the samples.

A single TMA core from an anonymised patient with corresponding IMC data was imaged with a pixel size of $10 \mu m \times 10 \mu m$ before upsampling by HyReS.

\subsubsection{Desorption Electro-Flow Focusing Ionization (DEFFI) imaging}
\textbf{Brain sections from Down Syndrome mouse model.} Ts65Dn mice (JAX stock no. 001924, B6EiC3Sn $a/A-Ts(1716)65Dn/J$) mice were purchased from Charles River, the mouse colony was amplified and maintainedat the Animal Facilities of the Barcelona Biomedical Research Park (PRBB, Barcelona, Spain, EU). All animal procedures met the guidelines of the European Community Directive 86/609/EEC and were approved by the Local Ethics Committee. Following breedings, the offspring were genotyped along the weaning stage, and segregated or co-housed, in sex-separated groups of 4-5 mice. Adult mice (4 months-old) were euthanized in a CO2 euthanasia chamber, immediately followed by brain hemispheres dissection, snap-freezing in liquid N2, and sample storage at -$80^{\circ} C$. All samples were cryosectioned at \SI{10}{\micro\metre} thickness using a Microm HM 550 Cryostat (Thermo Fisher Scientific Inc., Waltham, USA). Tissue sections were thaw-mounted onto SuperFrost Plus Glass slides (Thermo Fisher Scientific Inc., Waltham, USA). Tissue sections were stored at -$80^{\circ} C$ until further use. An enhanced desorption electrospary ionisation (DESI) imaging setup previously reported~\citep{Wu2022} was used for imaging of brain samples, using a commercially available DESI source (Waters Corporation) and previously optimised parameters~\citep{Tillner2017}.The sprayer utilised a TaperTip capillary measuring \SI{20}{\micro\metre} internally and \SI{363}{\micro\metre} externally, operating at a $75^{\circ}$ angle relative to the surface plane. It maintained distances of \SI{5}{\mm} from the sprayer to the inlet capillary and \SI{1}{\mm} from the sprayer to the surface, while applying a high voltage of 4.5 kV. A solvent mixture of methanol/water, 95:5 (v/v), was supplied by a nanoAcquity binary solvent manager (Waters, Wilmslow, UK) at a flow rate of \SI{1.5}{\micro\litre\per\min} and an inlet gas pressure of 5 bar (nitrogen). As the DEFFI sprayer was based on the same commercially available Waters DESI sprayer, the geometric configuration remained consistent. The final parameters for tissue imaging with DEFFI were established as follows: employing a TaperTip capillary with inner and outer diameters of \SI{20}{\micro\metre} and \SI{363}{\micro\metre} respectively, maintaining a capillary-to-orifice distance of \SI{100}{\micro\metre}, and utilizing an exit orifice diameter of \SI{150}{\micro\metre}. The solvent flow rate was adjusted to \SIrange[range-phrase={--}]{0.75}{1.5}{\micro\litre\per\min} of methanol/water, 95:5 (v/v), with an inlet gas pressure of 5 bar (nitrogen) and a high voltage of 4.5 kV. All mass spectrometric analysis was conducted in the negative ion mode within a spectral range of 50-1200 m/z. All Down Syndrome mouse brain sections were imaged with a pixel size of \SI{75}{\micro\metre}$\times$\SI{75}{\micro\metre} before upsampling by HyReS.\\

\textbf{Spinal cord sections from EAE mouse model.} Spinal cord sections of both control and acute EAE mice were imaged using the DEFFI system with a pixel size of \SI{70}{\micro\metre}$\times$\SI{70}{\micro\metre} for the study reported in~\citep{Peruzzotti-Jametti2024}. The same dataset was downsampled during preprocessing and then restored by HyReS to evaluate its effect on the biological information extracted.

\subsection{Imaging Mass Cytometry (IMC) of FFPE TMA Samples}
Formalin‑fixed, paraffin‑embedded (FFPE) tissue sections (4–10 µm) adjacent to those from Section \ref{subsec:reims} mounted on Superfrost slides were baked at 60 °C for 60 min in a heat‑resistant slide rack prior to staining. Slides were dewaxed twice in xylene (10 min each) and rehydrated through a graded ethanol series (100\%, 95\%, 80\%, 70\%; 5 min each), followed by a 5 min wash in double distilled water ($ddH_2O$) with gentle agitation (25–35 rpm) in Coplin jars. Antigen retrieval was performed in Target Retrieval Solution, pH 9 (Agilent Dako S2367) diluted to 1×, pre‑equilibrated at 96 °C in a water bath; slides were incubated in preheated retrieval buffer for 30 min with loose lids, then cooled on the bench for 20–30 min and washed in Maxpar water (2 × 10 min) and Maxpar DPBS (1 × 10 min) with gentle agitation. After gently blotting excess buffer, tissue areas were circled with a PAP pen and blocked for 45 min at room temperature in a humid chamber with blocking buffer (3\% BSA in DPBS supplemented with horse serum), prepared fresh from a 10\% BSA stock stored at $-20^{\circ}$C. For staining, a metal‑conjugated antibody cocktail was prepared at assay‑specific concentrations in Maxpar PBS containing 0.5\% BSA; individual antibodies were centrifuged at 13,000 × g for 2 min and supernatants were used to minimise aggregates, and the cocktail was kept on ice and applied within 1–2 h of preparation. After draining (without washing) the blocking buffer, sections were incubated with the antibody cocktail overnight at 4 °C in a humidified chamber. The next day, slides were washed twice for 8 min in wash buffer (0.2\% Triton‑X‑100 in PBS) and twice for 8 min in Maxpar PBS/DPBS, all with gentle orbital shaking. Nuclear staining was then performed with Cell‑ID Intercalator‑Ir (Fluidigm) diluted 1:400 in Maxpar PBS/DPBS (300–500 µL per section, approximately 20 $mm^2$) for 30 min at room temperature in a humid chamber, followed by a 5 min wash in Maxpar water on an orbital shaker. Finally, slides were air‑dried for at least 20 min at room temperature before IMC acquisition by Hyperion Imaging System (Standard Biotools (formerly Fluidigm)).

\subsection{Optical Imaging}
To generate the reference images for pathological assessment, consecutive slides obtained from the same sample sets (human breast, EAE mouse and FFPE TMA) used for MSI were H\&E-stained and scanned with a digital slide scanner (NanoZoomer2.0-HT, Hamamatsu City, Japan). A high-resolution objective (40$\times$) was used for all slides. After an initial autofocus procedure to identify the optimal focal positions of the cores, each slide was imaged within 10–20 min depending on the size of the effective ROI.

\subsection{Data Preprocessing}\label{sec:data_process}
Training data are first filtered to remove extremely low SNR images. The SNR is calculated by the mean squared error (MSE) in Equation~\ref{eq: MSE} between the image and pure black background, where, $D$ represents the matrix of the image and $B$ represents the matrix of the black background, $m$ represents the numbers of rows and $i$ represents the index of that row of the image. $n$ represents the number of columns and j represents the index of that column of the image. The background matrix only consists of zeros and therefore is discarded in the final equation.
\begin{equation}\label{eq: MSE}
M S E=\frac{1}{m n} \sum_{i=0}^{m-1} \sum_{j=0}^{n-1}(D(i, j)-B(i, j))^{2} = \frac{1}{m n} \sum_{i=0}^{m-1} \sum_{j=0}^{n-1}D(i, j)^{2}
\end{equation}

To enable FRCGAN training with only one set of images, the LR counterparts are synthetically generated from the original input dataset. For the restoration tasks attempted in this study with both synthetic and experimental data, first the images are cropped on both axes to be divisible by the scaling factor, in this case 4. The dynamic range of the images is normalised by the maximum and minimum intensities of the original image (which are saved separately and used to restore the dynamic range post-restoration) and then replicated three times to augment the RGB channels respectively. Next these images are downsampled using the standard bicubic kernel to create the LR counterparts that are 4 times smaller. To account for noise in the imaging system, we select a number of images randomly from the dataset and add additive noises (approximated to be Gaussian\citep{Xiang2023}) to both their fore- and backgrounds.


\subsection{SISR Network Architecture and Training}\label{sec:hyperselect}
HyReS utilises a SOTA GAN-based architecture for SISR (Real-ESRGAN)~\citep{Wang2021} that has achieved photo-realistic restoration and upsampling in natural images, which consists of a Generator and a Discriminator model. The Generator restores and upscales each LR input image and the resulting HR output image is then assessed by the Discriminator. The two models function hand-in-hand to iteratively improve the quality of the images generated by optimising some loss function. While common GAN-based SISR algorithms, including the real-ESRGAN assess the image quality based on some pixel-wise metric (e.g., L1 \& L2 norm), we have previously observed that they do not reflect the underlying imaging performance~\citep{Xiang2023} but rather based on some subjective perception of relative image quality. 

The limiting resolution of an imaging system is typically defined instrumentally and fundamentally related to the amount of information obtainable~\citep{Ober2004}. Hyperspectral imaging is no different, as we have previously reported a linear imaging model for MSI~\citep{Metodiev2021}:
\begin{equation}\label{model}
\mathrm{Detected~Signal} = \mathrm{True~Signal} \otimes \mathrm{Reponse} + \mathrm{Noise}
\end{equation}
where the acquired image can be seen as undergoing a blurring process, defined by a system response function, also known as the point spread function (PSF). The blurred image is then assumed to be further corrupted by additive noise. The task of a physically meaningful SISR model is thus to deconvolve and/or improve the PSF while minimising the noise simultaneously based on the input image alone.

To achieve this objective, PSF-relevant improvement in resolution, we have implemented two main modifications. Firstly, for all image restoration tasks, the Real-ESRGAN is fine-tuned in a self-supervised fashion with input MSI data that have been augmented into corresponding LR and HR sets with data dependent noise estimation (Section \ref{sec:data_process}). Additionally, to train models that are relevant in the scope of the physical imaging model (Sec.~\ref{sec:evaluation}), we have implemented a novel loss. This new loss ($\mathcal{L}_{\mathrm{FRC}}$) is based on Fourier Ring Correlation (FRC)~\citep{Koho2019}, 
\begin{equation}\label{eq:frc}
\mathrm{FRC} \left(r\right)=\frac{\sum_{r} F_1(r) \cdot F_2(r)^*}{\sqrt{\sum_{r} F_1^2(r) \cdot \sum_{r} F_2^2(r)}}
\end{equation}
where the $r$ is the frequency bin and $F_i$ is the Fourier transform of image $i$.
This effectively estimates the resolution of each image in frequency space with the assumption of a symmetric Gaussian PSF, in line with our previous derivation for MSI~\citep{Metodiev2021}. With the additional consideration of noise, this ensures relative improvement in terms of highest spatial frequency cut-off between input and output images, whereby:
 \begin{equation}
    \mathcal{L}_{\mathrm{FRC}} = 1-\sum_r \mathrm{FRC}(r)
\end{equation}
The new FRC loss has been checked to be differentiable and is therefore a well-defined loss. We have thus named this strategy FRCGAN. $\mathcal{L}_{\mathrm{FRC}}$ is minimised in the Generator of the architecture for all applications reported, while the Discriminator is still optimised via the Cross Entropy loss due to constraints in computational intensity. However, we expect that fast, probabilistic implementation into the Discriminator part of the network is feasible and may yield even better performance.

During training of the FRCGAN models, the Adam optimiser was used which yielded the following hyperparameter combination that produced the best upsampling performance: number of epochs = $100$, batch size= $8$, generator loss = $\mathcal{L}_{\mathrm{FRC}}$, discriminator loss = $\texttt{Cross Entropy}$, discriminator initial weight = $0$ and patch size = $50$.

\subsection{Evaluation Methods}\label{sec:evaluation}
We implement a combination of computational metrics to assess the restored images and the performance of our deep learning SISR models, even in the absence of ground truth images.

\subsubsection{Image Quality Metrics}

\textbf{Referenceless metrics.}
Although human observers can subjectively examine the restored images, image quality assessment (IQA) metrics are used in this work to quantitatively evaluate the performances of different SR models. The most common IQA metrics are full-reference metrics (comparing the test image against the reference image), for example, peak-signal-to-noise ratio (PSNR) and structural similarity index measure (SSIM). The PSNR calculates the image quality using a transform of the mean square error (MSE), and the SSIM is designed as a measure of structure, luminance, and contrast rather than the traditional error summation~\citep{Hore2010}. Most importantly, these are less suited for SISR models as the ground truth reference is not always available. There are also no-reference IQA metrics available (only using the test image for evaluation), for instance, blind/referenceless image spatial quality evaluator (BRISQUE) 
~\citep{Mittal2012a} and perception based image quality evaluator (PIQE)~\citep{Venkatanath2015}, for better estimating the perceptual quality of predicted images. BRISEQUE is an opinion-aware metric, and it requires a pre-trained model. The training images have known distortions, like artifacts, blurring and noise, and they have been marked with subjective quality scores. In contrast, PIQE is an opinion-unaware and unsupervised metric, which does not need any pre-training. PIQE can evaluate images with arbitrary distortion. 

The two referenceless metrics may disagree if the pretrained data set for BRISQUE diverge from the data set of interest thus results in difficult interpretation. We propose a new IQA metric CRISQUE, constrained referenceless image spatial quality evaluator, as a combination of BRISQUE and PIQE:
\begin{equation}
    \mathrm{CRISQUE} = \left(1- 2\times \frac{1}{\frac{1}{\mathrm{BRISQUE}}+\frac{1}{\mathrm{PIQE}}}\right)\times 100\%
\end{equation}
The fraction that takes both BRISQUE and PIQE into account is analogous to the resistance of two parallel resistors. This formulation ensures that CRISQUE is bounded between 0 and 100\% and can be easily interpreted as it increases as the image quality improves. Most importantly, when comparing SISR restored images with known reference, if BRISQUE increases (worse image quality), as long as PIQE decreases (better image quality) by equal or larger amount, the total score by CRISQUE still increases (better image quality).

\subsubsection{Image Resolution}
 While the standardisation of resolution remains an ongoing topic of discussion, we have previously presented a simple metric in the form of Fourier ring correlation~\citep{Xiang2023} that can estimate the PSF from a single image (Equation \ref{eq:frc}) and hence its estimated resolution. In addition to providing a quantifiable measure for image resolution comparison, this metric is also now embedded in our image restoration algorithms as a novel loss function (Section~\ref{sec:hyperselect}).
While two images are needed conventionally to compute FRC, here we adopt an approach that sub-divides each image into two images for FRC calculation~\citep{Koho2019} with a noise threshold determined information theoretically with the half-bit criterion\citep{VanHeel2005}. The half-bit threshold was also empirically chosen rather than the fixed 1/7 proposed by Koho et al., as it allowed for reasonable SNR in each half-image and produced the most consistent resolution values for our MSI datasets. It is possible that this subsampling approach may lead to undersampling and hence spurious resolution estimates, for which we have implemented a goodness-of-fit (GOF) check when computing the final cut-off spatial frequency by Gaussian curve fitting. Any value corresponding to a GOF $\le$ 0.75 is rejected and excluded from the results. 

\subsubsection{Difference Point Spread Function}
To generalise our physical imaging model and benchmark the performance of FRCGAN against the state-of-the-art, we derive a new metric that reflects the relative improvement to the imaging PSF. According to our proposed model (Eq.~\ref{model}), any raw image (\emph{I}) of the object (\emph{O}) in a hyperspectral dataset can be described by:
\begin{equation}
I = O \otimes PSF + N
\end{equation}
After applying an image restoration operation via FRCGAN or otherwise, we expect a direct impact on the PSF and noise (N) such that:
\begin{equation}
I' = O \otimes PSF' + N'
\end{equation}
where the restored image \emph{I'} is now convolved with a new system response (\emph{PSF'}) and noise level (\emph{N'}). To quantify this change, we first Fourier transform both the original and restored image and then take the ratio, which allows the use of well-known Fourier relations to cancel out the object function such that:
\begin{equation}
\frac{\hat{I}}{\hat{I'}} = \frac{M}{M'}+const
\end{equation}
where the Fourier transform operation is denoted by $\hat{}$. This gives rise to the Fourier-domain PSF, which is typically defined as the modulus transfer function (M) of the system and a constant term based on the assumption of white noise.

As we have previously reported a Gaussian PSF model in terms of MSI~\citep{Metodiev2021}, a new metric that measures the relative change in the imaging performance can be defined by taking the inverse Fourier transform:
\begin{equation}
PSF_d = F^{-1}(\frac{\hat{I}}{\hat{I'}})
\end{equation}

We name this metric the difference point spread function and use it to visualise the effect of HyReS and other restoration processes in a physically meaningful manner.

\subsection{Image Analysis}\label{sec:image analysis}
\subsubsection{Segmentation}
UMAP is a nonlinear dimensionality reduction technique that projects the high-dimensional data to a low-dimensional space while retaining the overall structures of the original data. For synthetically downsampled breast tissue data (Section \ref{section: bc}), 2D UMAP embeddings were generated for spatial clustering with the following hyperparameters, which have been empirically opitimised: (initialisation = spectral embedding of fuzzy 1-skeleton \citep{McInnes2018}, dimension = 2, number of neighbours = 20, distance metric = cosine, minimum distance between embedded points = 0.2). The resultant embeddings were then clustered using HDBSCAN, whose hyperparameters were also empirically optimised to best yield a pre-defined number of clusters for all analyses: (minimum sample size = 100, minium cluster size = 300, cluster selection method = Excess of Mass).

For the segmentation of the integrated multi-omic profiles of FFPE TMA core (Section \ref{section:ffpe}, accurate UMAP embeddings were first generated (initialisation = PCA, dimension = 5, number of neighbours = 50, distance metric = cosine, minimum distance = 0.0) to achieve a trustworthiness score. For explorative spatial clustering, a K-means method was chosen with k=4, pre-determined by finding the best-fit with the initial embeddings.
\subsubsection{Classification}
We also assessed if the restored images retain the biological features of the original images in terms of classification tasks. The regions of interests were selected manually from the original and upscaled bicubically for application in images that underwent pixel super-resolution (Section \ref{section:DS}). The pixel intensities for different (m/z) channels (aligned to 100 ppm) were then taken to train support vector machine (SVM) models, which were then evaluated by cross-validation (10-fold) and/or testing on independent data. The classification performance was generally evaluated by weighted accuracy~\citep{Mosley2013}. The whole spectral range was used for model training in the first iteration, to maximise model interpretability and the robustness of the model, we then performed recursive feature elimination (RFE) to select the only significant classification features which were used to train the optimised models. The hyperparameters in RFE were optimised through a grid search to determine the minimum features to select. 
For the classification of trisomy (Section \ref{section:DS}), we applied the optimised trained models first on whole brain tissue images (including regions unseen in training) that they were trained on and then independent brain images. A pixelwise probability was computed by the model as result in each case to generate heatmaps,
which was then used to label the whole tissue to be either trisomic or normal. This was assigned according to a percentage pixel threshold e.g., 50\% of all pixels pertaining to a tissue region in order for it to be classified. The threshold was adaptively optimised for each slide image (4 brains per slide) to yield the highest tissue-wise classification performance. For the whole tissue classification by GT and HyReS-enhanced images, the thresholds were set to be $(10\%, 10\%, 30\%, 20\%)$ and $(10\%, 10\%, 30\%, 30\%)$ respectively. For the independent test, the thresholds were $10\%$ (GT) and $20\%$ (HyReS).

\subsubsection{Imaging-based single cell multi-omics}
To perform multi-modal integration of FFPE TMA data obtained from various spatial omics and imaging techniques, the pixel resolutions were first universally harmonised to \SI{2.5}{\micro\metre}. To do so, the MSI data was HyReS upsampled 4-fold from its native resolution of \SI{10}{\micro\metre} and the adjacent IMC as well as brightfield images were downsampled from \SI{1}{\micro\metre} sampling rate while retaining subcellular resolution. Rigid affine transformations were then performed to maximise the spatial overlap between modalities using Mander's MOC coefficient \citep{Aaron2018}.
For downstream analysis, a 200 x 200 region-of-interest captured by all modalities was focused on. By first performing co-localisation analysis, the multi-omic features were trimmed to 29 protein and 87 metabolite targets which are spatially correalted with a Spearman coefficient $\geq 0.1$, corresponding p-value $\leq 0.05$ and MOC $\geq 0.5$. Before integration, variance stabilisation and empirical scaling was also carried out to equalise the dynamic ranges from different detectors, by:
\begin{equation}
    D' = \hat{Z}(log1p(D))
\end{equation}
where $\hat{Z}$ is the Z-score transformation, resulting in a scaled, multi-omic data matrix of dimension 40000 x 116.

To map the obtained segments to individual cells detected by IMC, the Ir193 channel was used, which is a common DNA intercalator and a specific single-cell marker that rules out debris and doublets~\citep{Iyer2022}. Cell mask was generated by thresholding, and in combination with individual image segment masks give the approximate number of cells per region. To assign cell types in these regions, the following cell types were identified with their corresponding core positive markers: tumour/epithelial cells (PANCK, Nd148), fibroblasts -- both cancer-associated and stromal (aSMA, Pr141), endothelial cells (CD31, Eu151), T cells (CD3, Er170), B cells (CD20, Dy121) and myeloid cells -- including those that are 'myeloid-like', such as M2‑ or repair‑associated macrophages (CD68, Tb159). These representative images are convolved with the cell mask per region, and cells that are detected above an intensity threshold are counted in each case. By summing all cell types and comparing with the pre-computed total number of cells, it was estimated that the accuracy of this approach is within $\pm 10\%$, still allowing for meaningful biological interpretation at a single-cell level.

\section*{Acknowledgements}
The authors thank C.K.L.Seow and D.Stoddart for the insightful discussions. We thank M.Wang for helping and curating the associated Github repository. Y.X, Z.T, N.S, J.M would like to thank the financial support from the CRUK Rosetta Grand Challenge (A25045, A24034, A25043, A25038) and BBSRC (BB/X004082/1) during this research.  R.T.M was supported by the Engineering and Physical Sciences Research Council (EPSRC) (EP/W029251/1). Z.L also acknowledges support by the EPSRC under grant EP/N014529/1 funding the EPSRC Centre for Mathematics of Precision Healthcare at Imperial College London. L.P.J was supported by funding from the Wellcome Trust Clinical Research Career Development Fellowship (G105713), Evelyn Trust Grant under the project reference 24/08 Med-24-2313, 
National MS Society Grant (RFA-2203-39318) and Medical Research Council (MRC) Clinician Scientist Fellowship (MR/Z50659X/1). M.E.G.S was supported by a School of Clinical Medicine Cambridge Trust Scholarship and the MRC Doctoral Training Grant (RG86932). XA is supported by ISCIII, co-funded by European Regional Development Fund (ERDF), a way to build Europe (PI25/00568, PCI25/00071), ERDERA Program (JTC25/099), Fundació La Marató TV3 (Project N. 607/C/2022), María de Maeztu Unit of Excellence (CEX2021-001159), and FEDER (AI-2023-038).

\section*{Author contributions}
Y.X and Z.L contributed equally to this manuscript and share first authorship. Y.X conceived and initiated the research. Y.X, Z.L, B.C designed, trained and validated the DL-SISR methods. D.S, V.W, K.R, Y.W, R.B acquired experimental data. R.T.M provided access to the high-resolution imaging setup. M.E.G.S and L.P.J acquired data and contributed to analyses of the EAE mouse models and revised the manuscript draft. X.A provided the DS mouse models. N.S and J.M acquired data and contributed in analyses on FFPE TMA data. Y.X and Z.T supervised the project. The manuscript writing was led by Y.X and Z.L with contributions and editing from all authors.

\newpage
\newpage

\newpage

\newpage

\newpage

\newpage

\newpage

\newpage

\newpage

\bibliographystyle{unsrtnat}
\bibliography{HyReS}  

@article{Jones2019c,
author = {Jones, Emrys A. and Simon, Daniel and Karancsi, Tamas and Balog, Julia and Pringle, Steven D. and Takats, Zoltan},
doi = {10.1021/acs.analchem.9b01441},
issn = {0003-2700},
journal = {Analytical Chemistry},
month = {aug},
number = {15},
pages = {9784--9791},
title = {{Matrix Assisted Rapid Evaporative Ionization Mass Spectrometry}},
url = {https://pubs.acs.org/doi/10.1021/acs.analchem.9b01441},
volume = {91},
year = {2019}
}

@article{Weigert2018,
author = {Weigert, Martin and Schmidt, Uwe and Boothe, Tobias and M{\"{u}}ller, Andreas and Dibrov, Alexandr and Jain, Akanksha and Wilhelm, Benjamin and Schmidt, Deborah and Broaddus, Coleman and Culley, Si{\^{a}}n and Rocha-Martins, Mauricio and Segovia-Miranda, Fabi{\'{a}}n and Norden, Caren and Henriques, Ricardo and Zerial, Marino and Solimena, Michele and Rink, Jochen and Tomancak, Pavel and Royer, Loic and Jug, Florian and Myers, Eugene W.},
doi = {10.1038/s41592-018-0216-7},
issn = {15487105},
journal = {Nature Methods},
number = {12},
pages = {1090--1097},
pmid = {30478326},
publisher = {Springer US},
title = {{Content-aware image restoration: pushing the limits of fluorescence microscopy}},
url = {http://dx.doi.org/10.1038/s41592-018-0216-7},
volume = {15},
year = {2018}
}

@article{McInnes2018,
archivePrefix = {arXiv},
arxivId = {1802.03426},
author = {McInnes, Leland and Healy, John and Melville, James},
eprint = {1802.03426},
title = {{UMAP: Uniform Manifold Approximation and Projection for Dimension Reduction}},
url = {http://arxiv.org/abs/1802.03426},
year = {2018}
}

@article{Borlinghaus2016,
author = {Borlinghaus, Rolf Theodor and Kappel, Constantin},
doi = {10.1038/nmeth.f.392},
issn = {1548-7091},
journal = {Nature Methods},
number = {3},
pages = {i--iii},
publisher = {Nature Publishing Group},
title = {{HyVolution—the smart path to confocal super-resolution}},
volume = {13},
year = {2016}
}

@article{Peruzzotti-Jametti2021,
author = {Peruzzotti-Jametti, Luca and Willis, Cory M. and Hamel, Regan and Krzak, Grzegorz and Pluchino, Stefano},
doi = {10.3389/fimmu.2021.705920},
issn = {16643224},
journal = {Frontiers in Immunology},
number = {June},
pages = {1--16},
pmid = {34249016},
title = {{Metabolic Control of Smoldering Neuroinflammation}},
volume = {12},
year = {2021}
}

@article{Wang2021,
archivePrefix = {arXiv},
arxivId = {2107.10833},
author = {Wang, Xintao and Xie, Liangbin and Dong, Chao and Shan, Ying},
eprint = {2107.10833},
month = {jul},
title = {{Real-ESRGAN: Training Real-World Blind Super-Resolution with Pure Synthetic Data}},
url = {http://arxiv.org/abs/2107.10833},
year = {2021}
}

@article{Tipping2003,
author = {Tipping, Michael E. and Bishop, Christopher M.},
isbn = {0262025507},
issn = {10495258},
journal = {Advances in Neural Information Processing Systems},
title = {{Bayesian image super-resolution}},
year = {2003}
}

@article{Rachidi2007,
author = {Rachidi, Mohammed and Lopes, Carmela},
doi = {10.1016/j.neures.2007.08.007},
issn = {01680102},
journal = {Neuroscience Research},
month = {dec},
number = {4},
pages = {349--369},
title = {{Mental retardation in Down syndrome: From gene dosage imbalance to molecular and cellular mechanisms}},
url = {https://linkinghub.elsevier.com/retrieve/pii/S0168010207017464},
volume = {59},
year = {2007}
}

@article{Wang2019,
archivePrefix = {arXiv},
arxivId = {1809.00219},
author = {Wang, Xintao and Yu, Ke and Wu, Shixiang and Gu, Jinjin and Liu, Yihao and Dong, Chao and Qiao, Yu and Loy, Chen Change},
doi = {10.1007/978-3-030-11021-5_5},
eprint = {1809.00219},
isbn = {9783030110208},
issn = {16113349},
journal = {Lecture Notes in Computer Science (including subseries Lecture Notes in Artificial Intelligence and Lecture Notes in Bioinformatics)},
pages = {63--79},
title = {{ESRGAN: Enhanced super-resolution generative adversarial networks}},
volume = {11133 LNCS},
year = {2019}
}

@article{Li2013,
author = {Li, Shuzhao and Park, Youngja and Duraisingham, Sai and Strobel, Frederick H. and Khan, Nooruddin and Soltow, Quinlyn A. and Jones, Dean P. and Pulendran, Bali},
doi = {10.1371/journal.pcbi.1003123},
editor = {Ouzounis, Christos A.},
issn = {1553-7358},
journal = {PLoS Computational Biology},
month = {jul},
number = {7},
pages = {e1003123},
title = {{Predicting Network Activity from High Throughput Metabolomics}},
url = {https://dx.plos.org/10.1371/journal.pcbi.1003123},
volume = {9},
year = {2013}
}

@article{Pecze2020,
author = {Pecze, Laszlo and Randi, Elisa B. and Szabo, Csaba},
doi = {10.1186/s10020-020-00225-8},
issn = {1076-1551},
journal = {Molecular Medicine},
month = {dec},
number = {1},
pages = {102},
title = {{Meta-analysis of metabolites involved in bioenergetic pathways reveals a pseudohypoxic state in Down syndrome}},
volume = {26},
year = {2020}
}

@article{Caracausi2018,
author = {Caracausi, Maria and Ghini, Veronica and Locatelli, Chiara and Mericio, Martina and Piovesan, Allison and Antonaros, Francesca and Pelleri, Maria Chiara and Vitale, Lorenza and Vacca, Rosa Anna and Bedetti, Federica and Mimmi, Maria Chiara and Luchinat, Claudio and Turano, Paola and Strippoli, Pierluigi and Cocchi, Guido},
doi = {10.1038/s41598-018-20834-y},
issn = {2045-2322},
journal = {Scientific Reports},
month = {feb},
number = {1},
pages = {2977},
title = {{Plasma and urinary metabolomic profiles of Down syndrome correlate with alteration of mitochondrial metabolism}},
volume = {8},
year = {2018}
}

@article{Goodman2018,
author = {Goodman, Jarid and Packard, Mark G.},
doi = {10.1016/j.nlm.2018.02.028},
issn = {10747427},
journal = {Neurobiology of Learning and Memory},
month = {apr},
pages = {48--55},
title = {{The role of the dorsal striatum in extinction: A memory systems perspective}},
volume = {150},
year = {2018}
}

@article{Simon2025,
	title = {Subcellular-{Resolution} {Molecular} {Pathology} by {Laser} {Ablation}–{Rapid} {Evaporative} {Ionization} {Mass} {Spectrometry}},
	volume = {97},
	copyright = {https://creativecommons.org/licenses/by/4.0/},
	issn = {0003-2700, 1520-6882},
	url = {https://pubs.acs.org/doi/10.1021/acs.analchem.5c02013},
	doi = {10.1021/acs.analchem.5c02013},
	language = {en},
	number = {32},
	urldate = {2026-01-28},
	journal = {Analytical Chemistry},
	author = {Simon, Dániel and Horkovics-Kováts, Gabriel Stefan and Xiang, Yuchen and Battle, Ronan A. and Wang, Yu and Abda, Julia and Papanastasiou, Dimitris and Stavrakaki, Stefania Maneta and Ho, Hui-Yu and Wang, Haixing and Schäffer, Richard and Karancsi, Tamás and Mroz, Anna and Pap, István and Lagache, Laurine and Balog, Júlia and Fournier, Isabelle and Murray, Robert T. and Bunch, Josephine and Takáts, Zoltan},
	month = aug,
	year = {2025},
	pages = {17433--17443},
}

@article{Wu2022,
author = {Wu, Vincen and Tillner, Jocelyn and Jones, Emrys and McKenzie, James S. and Gurung, Dipa and Mroz, Anna and Poynter, Liam and Simon, Daniel and Grau, Cristina and Altafaj, Xavier and Dumas, Marc-Emmanuel and Gilmore, Ian and Bunch, Josephine and Takats, Zoltan},
doi = {10.1021/acs.analchem.2c00345},
issn = {0003-2700},
journal = {Analytical Chemistry},
month = {jul},
number = {28},
pages = {10035--10044},
title = {{High Resolution Ambient MS Imaging of Biological Samples by Desorption Electro-Flow Focussing Ionization}},
url = {https://pubs.acs.org/doi/10.1021/acs.analchem.2c00345},
volume = {94},
year = {2022}
}

@article{Cai2023,
author = {Cai, Shaoli and Lin, Jinxin and Li, Zhaolong and Liu, Songnian and Feng, Zhihua and Zhang, Yangfan and Zhang, Yanding and Huang, Jianzhong and Chen, Qi},
doi = {10.3389/fmicb.2023.1016872},
issn = {1664-302X},
journal = {Frontiers in Microbiology},
month = {feb},
title = {{Alterations in intestinal microbiota and metabolites in individuals with Down syndrome and their correlation with inflammation and behavior disorders in mice}},
volume = {14},
year = {2023}
}

@article{Venkatanath2015,
author = {Venkatanath, N. and Praneeth, D. and {Maruthi Chandrasekhar}, B. H. and Channappayya, Sumohana S. and Medasani, Swarup S.},
doi = {10.1109/NCC.2015.7084843},
isbn = {9781479966196},
journal = {2015 21st National Conference on Communications, NCC 2015},
title = {{Blind image quality evaluation using perception based features}},
year = {2015}
}

@article{Iveson2018,
author = {Iveson, Timothy J and Kerr, Rachel S and Saunders, Mark P and Cassidy, Jim and Hollander, Niels Henrik and Tabernero, Josep and Haydon, Andrew and Glimelius, Bengt and Harkin, Andrea and Allan, Karen and McQueen, John and Scudder, Claire and Boyd, Kathleen Anne and Briggs, Andrew and Waterston, Ashita and Medley, Louise and Wilson, Charles and Ellis, Richard and Essapen, Sharadah and Dhadda, Amandeep S and Harrison, Mark and Falk, Stephen and Raouf, Sherif and Rees, Charlotte and Olesen, Rene K and Propper, David and Bridgewater, John and Azzabi, Ashraf and Farrugia, David and Webb, Andrew and Cunningham, David and Hickish, Tamas and Weaver, Andrew and Gollins, Simon and Wasan, Harpreet S and Paul, James},
doi = {10.1016/S1470-2045(18)30093-7},
issn = {14702045},
journal = {The Lancet Oncology},
month = {apr},
number = {4},
pages = {562--578},
title = {{3 versus 6 months of adjuvant oxaliplatin-fluoropyrimidine combination therapy for colorectal cancer (SCOT): an international, randomised, phase 3, non-inferiority trial}},
url = {https://linkinghub.elsevier.com/retrieve/pii/S1470204518300937},
volume = {19},
year = {2018}
}

@article{Rueda2012,
author = {Rueda, Noem{\'{i}} and Fl{\'{o}}rez, Jes{\'{u}}s and Mart{\'{i}}nez-Cu{\'{e}}, Carmen},
doi = {10.1155/2012/584071},
issn = {2090-5904},
journal = {Neural Plasticity},
pages = {1--26},
title = {{Mouse Models of Down Syndrome as a Tool to Unravel the Causes of Mental Disabilities}},
url = {http://www.hindawi.com/journals/np/2012/584071/},
volume = {2012},
year = {2012}
}

@article{Kabakova2024,
author = {Kabakova, Irina and Zhang, Jitao and Xiang, Yuchen and Caponi, Silvia and Bilenca, Alberto and Guck, Jochen and Scarcelli, Giuliano},
doi = {10.1038/s43586-023-00286-z},
issn = {2662-8449},
journal = {Nature Reviews Methods Primers},
month = {feb},
number = {1},
pages = {8},
title = {{Brillouin microscopy}},
url = {https://www.nature.com/articles/s43586-023-00286-z},
volume = {4},
year = {2024}
}

@article{Metodiev2021,
author = {Metodiev, Martin D. and Steven, Rory T. and Loizeau, Xavier and Takats, Zoltan and Bunch, Josephine},
doi = {10.1021/acs.analchem.1c02470},
issn = {0003-2700},
journal = {Analytical Chemistry},
month = {nov},
number = {46},
pages = {15295--15305},
title = {{Modality Agnostic Model for Spatial Resolution in Mass Spectrometry Imaging: Application to MALDI MSI Data}},
url = {https://pubs.acs.org/doi/10.1021/acs.analchem.1c02470},
volume = {93},
year = {2021}
}

@article{Xue2019,
author = {Xue, Jinjuan and Bai, Yu and Liu, Huwei},
doi = {10.1016/j.trac.2019.115659},
issn = {18793142},
journal = {TrAC - Trends in Analytical Chemistry},
pages = {115659},
publisher = {Elsevier Ltd},
title = {{Recent advances in ambient mass spectrometry imaging}},
url = {https://doi.org/10.1016/j.trac.2019.115659},
volume = {120},
year = {2019}
}

@article{Worley2023,
author = {Worley, Gordon and Byeon, Seul Kee and Smith, P. Brian and Hart, Sarah J. and Young, Sarah P. and Pandey, Akhilesh and Kishnani, Priya S.},
doi = {10.1002/ajmg.a.63325},
issn = {1552-4825},
journal = {American Journal of Medical Genetics Part A},
month = {sep},
number = {9},
pages = {2300--2311},
title = {{An exploratory study of plasma ceramides in comorbidities in Down syndrome}},
volume = {191},
year = {2023}
}

@phdthesis{Mosley2013,
address = {Ames},
author = {Mosley, Lawrence},
doi = {10.31274/etd-180810-3375},
school = {Iowa State University, Digital Repository},
title = {{A balanced approach to the multi-class imbalance problem}},
url = {https://lib.dr.iastate.edu/etd/13537/},
year = {2013}
}

@article{Yoon2022,
author = {Yoon, Jonghee},
doi = {10.1007/s13206-021-00041-0},
issn = {1976-0280},
journal = {BioChip Journal},
month = {mar},
number = {1},
pages = {1--12},
title = {{Hyperspectral Imaging for Clinical Applications}},
url = {https://link.springer.com/10.1007/s13206-021-00041-0},
volume = {16},
year = {2022}
}

@inproceedings{EsterM1996,
author = {Ester, Martin and Kriegel, Hans-Peter and Sander, J{\"{o}}rg and Xu, Xiaowei},
booktitle = {Proceedings of the Second International Conference on Knowledge Discovery and Data Mining},
pages = {226--231},
publisher = {AAAI Press},
series = {KDD'96},
title = {{A Density-Based Algorithm for Discovering Clusters a Density-Based Algorithm for Discovering Clusters in Large Spatial Databases with Noise}},
year = {1996}
}

@article{Peruzzotti-Jametti2024,
author = {Peruzzotti-Jametti, L. and Willis, C. M. and Krzak, G. and Hamel, R. and Pirvan, L. and Ionescu, R.-B. and Reisz, J. A. and Prag, H. A. and Garcia-Segura, M. E. and Wu, V. and Xiang, Y. and Barlas, B. and Casey, A. M. and van den Bosch, A. M. R. and Nicaise, A. M. and Roth, L. and Bates, G. R. and Huang, H. and Prasad, P. and Vincent, A. E. and Frezza, C. and Viscomi, C. and Balmus, G. and Takats, Z. and Marioni, J. C. and D'Alessandro, A. and Murphy, M. P. and Mohorianu, I. and Pluchino, S.},
doi = {10.1038/s41586-024-07167-9},
issn = {0028-0836},
journal = {Nature},
month = {apr},
number = {8006},
pages = {195--203},
title = {{Mitochondrial complex I activity in microglia sustains neuroinflammation}},
url = {https://www.nature.com/articles/s41586-024-07167-9},
volume = {628},
year = {2024}
}

@incollection{Singh2020,
author = {Singh, Prachi and Pandey, Prem Chandra and Petropoulos, George P. and Pavlides, Andrew and Srivastava, Prashant K. and Koutsias, Nikos and Deng, Khidir Abdala Kwal and Bao, Yangson},
booktitle = {Hyperspectral Remote Sensing},
doi = {10.1016/B978-0-08-102894-0.00009-7},
pages = {121--146},
publisher = {Elsevier},
title = {{Hyperspectral remote sensing in precision agriculture: present status, challenges, and future trends}},
url = {https://linkinghub.elsevier.com/retrieve/pii/B9780081028940000097},
year = {2020}
}

@article{Iyer2022,
author = {Iyer, Akshay and Hamers, Anouk A.J. and Pillai, Asha B.},
doi = {10.3389/fimmu.2022.815828},
issn = {16643224},
journal = {Frontiers in Immunology},
number = {April},
pages = {14--17},
pmid = {35493491},
title = {{CyTOF{\textregistered} for the Masses}},
volume = {13},
year = {2022}
}

@article{Ober2004,
author = {Ober, Raimund J. and Ram, Sripad and Ward, E. Sally},
doi = {10.1016/S0006-3495(04)74193-4},
issn = {00063495},
journal = {Biophysical Journal},
number = {2},
pages = {1185--1200},
publisher = {Elsevier},
title = {{Localization Accuracy in Single-Molecule Microscopy}},
url = {http://dx.doi.org/10.1016/S0006-3495(04)74193-4},
volume = {86},
year = {2004}
}

@article{Abdo2019,
author = {Abdo, Mohammad and Badilita, Vlad and Korvink, Jan},
doi = {10.1364/oe.27.020290},
issn = {1094-4087},
journal = {Optics Express},
number = {15},
pages = {20290},
title = {{Spatial scanning hyperspectral imaging combining a rotating slit with a Dove prism}},
volume = {27},
year = {2019}
}

@article{Hu2022,
author = {Hu, Hang and Laskin, Julia},
doi = {10.1002/advs.202203339},
issn = {21983844},
journal = {Advanced Science},
number = {34},
pages = {1--20},
pmid = {36253139},
title = {{Emerging Computational Methods in Mass Spectrometry Imaging}},
volume = {9},
year = {2022}
}

@article{Fabelo2018,
author = {Fabelo, Himar and Ortega, Samuel and Ravi, Daniele and Kiran, B. Ravi and Sosa, Coralia and Bulters, Diederik and Callic{\'{o}}, Gustavo M. and Bulstrode, Harry and Szolna, Adam and Pi{\~{n}}eiro, Juan F. and Kabwama, Silvester and Madro{\~{n}}al, Daniel and Lazcano, Raquel and J-O'Shanahan, Aruma and Bisshopp, Sara and Hern{\'{a}}ndez, Mar{\'{i}}a and B{\'{a}}ez, Abelardo and Yang, Guang-Zhong and Stanciulescu, Bogdan and Salvador, Rub{\'{e}}n and Ju{\'{a}}rez, Eduardo and Sarmiento, Roberto},
doi = {10.1371/journal.pone.0193721},
editor = {Fred, A. Lenin},
isbn = {1111111111},
issn = {1932-6203},
journal = {PLOS ONE},
month = {mar},
number = {3},
pages = {e0193721},
title = {{Spatio-spectral classification of hyperspectral images for brain cancer detection during surgical operations}},
url = {https://dx.plos.org/10.1371/journal.pone.0193721},
volume = {13},
year = {2018}
}

@article{Tillner2017,
author = {Tillner, Jocelyn and Wu, Vincen and Jones, Emrys A. and Pringle, Steven D. and Karancsi, Tamas and Dannhorn, Andreas and Veselkov, Kirill and McKenzie, James S. and Takats, Zoltan},
doi = {10.1007/s13361-017-1714-z},
isbn = {1336101717},
issn = {18791123},
journal = {Journal of the American Society for Mass Spectrometry},
number = {10},
pages = {2090--2098},
title = {{Faster, More Reproducible DESI-MS for Biological Tissue Imaging}},
volume = {28},
year = {2017}
}

@article{Battle2023,
author = {Battle, Ronan A. and Chandran, Anita M. and Runcorn, Timothy H. and Mussot, Arnaud and Kudlinski, Alexandre and Murray, Robert T. and {Roy Taylor}, J.},
doi = {10.1364/OL.476754},
issn = {0146-9592},
journal = {Optics Letters},
month = {jan},
number = {2},
pages = {387},
title = {{Mid-infrared difference-frequency generation directly pumped by a fiber four-wave mixing source}},
url = {https://opg.optica.org/abstract.cfm?URI=ol-48-2-387},
volume = {48},
year = {2023}
}

@article{VanDePlas2015,
author = {{Van de Plas}, Raf and Yang, Junhai and Spraggins, Jeffrey and Caprioli, Richard M.},
doi = {10.1038/nmeth.3296},
issn = {1548-7091},
journal = {Nature Methods},
month = {apr},
number = {4},
pages = {366--372},
pmid = {25707028},
title = {{Image fusion of mass spectrometry and microscopy: a multimodality paradigm for molecular tissue mapping}},
url = {http://www.nature.com/articles/nmeth.3296},
volume = {12},
year = {2015}
}

@article{Wiseman2006,
author = {Wiseman, Justin M. and Ifa, Demian R. and Song, Qingyu and Cooks, R. Graham},
doi = {10.1002/anie.200602449},
issn = {14337851},
journal = {Angewandte Chemie International Edition},
month = {nov},
number = {43},
pages = {7188--7192},
title = {{Tissue Imaging at Atmospheric Pressure Using Desorption Electrospray Ionization (DESI) Mass Spectrometry}},
url = {https://onlinelibrary.wiley.com/doi/10.1002/anie.200602449},
volume = {45},
year = {2006}
}

@misc{Dierssen2020,
author = {Dierssen, Mara and Fructuoso, Marta and {Mart{\'{i}}nez de Lagr{\'{a}}n}, Mar{\'{i}}a and Perluigi, Marzia and Barone, Eugenio},
booktitle = {Frontiers in Neuroscience},
doi = {10.3389/fnins.2020.00670},
issn = {1662453X},
month = {jul},
publisher = {Frontiers Media S.A.},
title = {{Down Syndrome Is a Metabolic Disease: Altered Insulin Signaling Mediates Peripheral and Brain Dysfunctions}},
volume = {14},
year = {2020}
}

@article{Koho2019,
author = {Koho, Sami and Tortarolo, Giorgio and Castello, Marco and Deguchi, Takahiro and Diaspro, Alberto and Vicidomini, Giuseppe},
doi = {10.1038/s41467-019-11024-z},
issn = {20411723},
journal = {Nature Communications},
number = {1},
pmid = {31308370},
title = {{Fourier ring correlation simplifies image restoration in fluorescence microscopy}},
volume = {10},
year = {2019}
}

@article{Xiang2023,
author = {Xiang, Yuchen and Metodiev, Martin and Wang, Meiqi and Cao, Boxuan and Bunch, Josephine and Takats, Zoltan},
doi = {10.3390/metabo13050669},
issn = {2218-1989},
journal = {Metabolites},
month = {may},
number = {5},
pages = {669},
title = {{Enhancement of Ambient Mass Spectrometry Imaging Data by Image Restoration}},
url = {https://www.mdpi.com/2218-1989/13/5/669},
volume = {13},
year = {2023}
}

@article{Jones2019a,
author = {Jones, Robin R. and Hooper, David C. and Zhang, Liwu and Wolverson, Daniel and Valev, Ventsislav K.},
doi = {10.1186/s11671-019-3039-2},
issn = {1931-7573},
journal = {Nanoscale Research Letters},
month = {dec},
number = {1},
pages = {231},
title = {{Raman Techniques: Fundamentals and Frontiers}},
url = {https://nanoscalereslett.springeropen.com/articles/10.1186/s11671-019-3039-2},
volume = {14},
year = {2019}
}

@article{Huang2022,
author = {Huang, Gao and Liu, Zhuang and Pleiss, Geoff and van der Maaten, Laurens and Weinberger, Kilian Q.},
doi = {10.1109/TPAMI.2019.2918284},
issn = {0162-8828},
journal = {IEEE Transactions on Pattern Analysis and Machine Intelligence},
month = {dec},
number = {12},
pages = {8704--8716},
title = {{Convolutional Networks with Dense Connectivity}},
url = {https://ieeexplore.ieee.org/document/8721151/},
volume = {44},
year = {2022}
}

@article{Mittal2012a,
author = {Mittal, A. and Moorthy, A. K. and Bovik, A. C.},
doi = {10.1109/TIP.2012.2214050},
issn = {1057-7149},
journal = {IEEE Transactions on Image Processing},
month = {dec},
number = {12},
pages = {4695--4708},
title = {{No-Reference Image Quality Assessment in the Spatial Domain}},
url = {http://ieeexplore.ieee.org/document/6272356/},
volume = {21},
year = {2012}
}

@article{Ledig2016,
archivePrefix = {arXiv},
arxivId = {1609.04802},
author = {Ledig, Christian and Theis, Lucas and Huszar, Ferenc and Caballero, Jose and Cunningham, Andrew and Acosta, Alejandro and Aitken, Andrew and Tejani, Alykhan and Totz, Johannes and Wang, Zehan and Shi, Wenzhe},
eprint = {1609.04802},
month = {sep},
title = {{Photo-Realistic Single Image Super-Resolution Using a Generative Adversarial Network}},
url = {http://arxiv.org/abs/1609.04802},
year = {2016}
}

@article{Xiao2020,
author = {Xiao, Yipo and Deng, Jiewei and Yao, Yao and Fang, Ling and Yang, Yunyun and Luan, Tiangang},
doi = {10.1016/j.aca.2020.01.052},
issn = {18734324},
journal = {Analytica Chimica Acta},
pages = {74--88},
pmid = {32408956},
publisher = {Elsevier Ltd},
title = {{Recent advances of ambient mass spectrometry imaging for biological tissues: A review}},
url = {https://doi.org/10.1016/j.aca.2020.01.052},
volume = {1117},
year = {2020}
}

@article{Passeri2023,
author = {Passeri, A.A. and {Di Michele}, A. and Neri, I. and Cottone, F. and Fioretto, D. and Mattarelli, M. and Caponi, S.},
doi = {10.1016/j.bioadv.2023.213341},
issn = {27729508},
journal = {Biomaterials Advances},
month = {apr},
pages = {213341},
title = {{Size and environment: The effect of phonon localization on micro-Brillouin imaging}},
url = {https://linkinghub.elsevier.com/retrieve/pii/S277295082300064X},
volume = {147},
year = {2023}
}

@article{Qiao2021,
author = {Qiao, Chang and Li, Di and Guo, Yuting and Liu, Chong and Jiang, Tao and Dai, Qionghai and Li, Dong},
doi = {10.1038/s41592-020-01048-5},
issn = {1548-7091},
journal = {Nature Methods},
month = {feb},
number = {2},
pages = {194--202},
pmid = {33479522},
publisher = {Nature Research},
title = {{Evaluation and development of deep neural networks for image super-resolution in optical microscopy}},
url = {http://www.nature.com/articles/s41592-020-01048-5},
volume = {18},
year = {2021}
}

@article{Liao2023,
author = {Liao, Tiepeng and Ren, Zihao and Chai, Zhaoliang and Yuan, Man and Miao, Chenjian and Li, Junjie and Chen, Qi and Li, Zhilin and Wang, Ziyi and Yi, Lin and Ge, Siyuan and Qian, Wenwei and Shen, Longfeng and Wang, Zilei and Xiong, Wei and Zhu, Hongying},
doi = {10.1038/s42256-023-00677-7},
issn = {25225839},
journal = {Nature Machine Intelligence},
number = {6},
pages = {656--668},
publisher = {Springer US},
title = {{A super-resolution strategy for mass spectrometry imaging via transfer learning}},
volume = {5},
year = {2023}
}

@inproceedings{Hore2010,
author = {Hore, Alain and Ziou, Djemel},
booktitle = {2010 20th International Conference on Pattern Recognition},
doi = {10.1109/ICPR.2010.579},
isbn = {978-1-4244-7542-1},
month = {aug},
pages = {2366--2369},
publisher = {IEEE},
title = {{Image Quality Metrics: PSNR vs. SSIM}},
url = {http://ieeexplore.ieee.org/document/5596999/},
year = {2010}
}

@inproceedings{Liu2013,
author = {Liu, Jing and Gan, Zongliang and Zhu, Xiuchang},
doi = {10.2991/icmt-13.2013.57},
title = {{Directional Bicubic Interpolation — A New Method of Image Super-Resolution}},
url = {https://www.atlantis-press.com/article/10409},
year = {2013}
}

@article{andac-aktas_immunological_2025,
	title = {Immunological landscape of colorectal cancer: tumor microenvironment, cellular players and immunotherapeutic opportunities},
	volume = {12},
	issn = {2296-889X},
	shorttitle = {Immunological landscape of colorectal cancer},
	url = {https://pmc.ncbi.nlm.nih.gov/articles/PMC12611834/},
	doi = {10.3389/fmolb.2025.1687556},
	urldate = {2026-04-03},
	journal = {Frontiers in Molecular Biosciences},
	author = {Andac-Aktas, Adile Buse and Calibasi-Kocal, Gizem},
	month = oct,
	year = {2025},
	pages = {1687556},
}

@article{liu_survey_2021,
	title = {A survey on applications of deep learning in microscopy image analysis},
	volume = {134},
	issn = {0010-4825},
	url = {https://www.sciencedirect.com/science/article/pii/S0010482521003176},
	doi = {10.1016/j.compbiomed.2021.104523},
	urldate = {2026-04-03},
	journal = {Computers in Biology and Medicine},
	author = {Liu, Zhichao and Jin, Luhong and Chen, Jincheng and Fang, Qiuyu and Ablameyko, Sergey and Yin, Zhaozheng and Xu, Yingke},
	month = jul,
	year = {2021},
	pages = {104523},
}

@article{Aaron2018,
	title = {Image co-localization - co-occurrence versus correlation},
	volume = {131},
	issn = {14779137},
	doi = {10.1242/jcs.211847},
	number = {3},
	journal = {Journal of cell science},
	author = {Aaron, Jesse S. and Taylor, Aaron B. and Chew, Teng Leong},
	year = {2018},
}

@article{VanHeel2005,
	title = {Fourier shell correlation threshold criteria},
	volume = {151},
	issn = {10478477},
	doi = {10.1016/j.jsb.2005.05.009},
	number = {3},
	journal = {Journal of Structural Biology},
	author = {Van Heel, Marin and Schatz, Michael},
	year = {2005},
	pages = {250--262},
}

@article{nunes_integration_2024,
	title = {Integration of mass cytometry and mass spectrometry imaging for spatially resolved single-cell metabolic profiling},
	volume = {21},
	copyright = {2024 The Author(s)},
	issn = {1548-7105},
	url = {https://www.nature.com/articles/s41592-024-02392-6},
	doi = {10.1038/s41592-024-02392-6},
	language = {en},
	number = {10},
	urldate = {2026-08-06},
	journal = {Nature Methods},
	publisher = {Nature Publishing Group},
	author = {Nunes, Joana B. and Ijsselsteijn, Marieke E. and Abdelaal, Tamim and Ursem, Rick and van der Ploeg, Manon and Giera, Martin and Everts, Bart and Mahfouz, Ahmed and Heijs, Bram and de Miranda, Noel F. C. C.},
	month = oct,
	year = {2024},
	pages = {1796--1800},
}

@article{potthoff_spatial_2025,
	title = {Spatial biology using single-cell mass spectrometry imaging and integrated microscopy},
	volume = {16},
	copyright = {2025 The Author(s)},
	issn = {2041-1723},
	url = {https://www.nature.com/articles/s41467-025-64603-8},
	doi = {10.1038/s41467-025-64603-8},
	language = {en},
	number = {1},
	urldate = {2026-08-06},
	journal = {Nature Communications},
	publisher = {Nature Publishing Group},
	author = {Potthoff, Alexander and Schwenzfeier, Jan and Niehaus, Marcel and Bessler, Sebastian and Hoffmann, Emily and Soehnlein, Oliver and Höhndorf, Jens and Dreisewerd, Klaus and Soltwisch, Jens},
	month = oct,
	year = {2025},
	pages = {9129},
}

\end{document}